\documentclass[sigconf,balance=false]{acmart}
\usepackage{popets}
\usepackage{pifont}
\usepackage{float}

\setcopyright{popets}
\copyrightyear{YYYY}

\acmYear{YYYY}
\acmVolume{YYYY}
\acmNumber{X}
\acmDOI{XXXXXXX.XXXXXXX}
\acmISBN{}
\acmConference{Proceedings on Privacy Enhancing Technologies}
\begin{document}

\title[Surveillance Disguised as Protection]{Surveillance Disguised as Protection: A Comparative Analysis of Sideloaded and In-Store Parental Control Apps}

\author{Eva-Maria Maier}
\affiliation{%
  \institution{St.\ Pölten UAS}
  \city{} 
  \state{} 
  \country{Austria} 
}
\email{eva-maria.maier@proton.me}

\author{Leonie Maria Tanczer}
\affiliation{%
  \institution{University College London}
  \city{}
  \country{United Kingdom}}
\email{l.tanczer@ucl.ac.uk}

\author{Lukas Daniel Klausner}
\affiliation{%
  \institution{St.\ Pölten UAS}
  \city{}
  \country{Austria}
}
\email{mail@l17r.eu}

\renewcommand{\shortauthors}{Maier, Tanczer and Klausner}

\begin{abstract}
    Parental control applications, software tools designed to manage and monitor children’s online activities, serve as essential safeguards for parents in the digital age. However, their usage has sparked concerns about security and privacy violations inherent in various child monitoring products.
    Sideloaded software (i.\,e.\ apps installed outside official app stores) poses an increased risk, as it is not bound by the regulations of trusted platforms.
    Despite this, the market of sideloaded parental control software has remained widely unexplored by the research community. 
    This paper examines 20 sideloaded parental control apps and compares them to 20 apps available on the Google Play Store.
    We base our analysis on privacy policies, Android package kit (APK) files, application behaviour, network traffic and application functionalities.
    Our findings reveal that sideloaded parental control apps fall short compared to their in-store counterparts, lacking specialised parental control features and safeguards against misuse while concealing themselves on the user's device. 
    Alarmingly, three apps transmitted sensitive data unencrypted, half lacked a privacy policy and 8 out of 20 were flagged for potential stalkerware indicators of compromise (IOC).
\end{abstract}


\maketitle

\section{Introduction}
        Parental control systems are increasingly used by parents\footnote{\ We use the term ``parent'' to refer to the adults entrusted with the responsibility of caregiving and guardianship, including parents \emph{stricto sensu}, legal guardians and other adults playing a central role in a child's upbringing.} to protect their children’s safety, security and privacy.\footnote{\ For the purpose of our paper, ``safety'' refers to protection against dangers and harm, ``security'' involves protecting children from online threats and ``privacy'' encompasses children's data protection.} 
        They can be classified into time, content and activity restricting as well as monitoring and tracking services~\cite{Zaman2016Parental}. They often provide functionalities such as logging and limiting screentime, browser history monitoring and blocking of apps~\cite{Wang2021Protection}. According to a report by Internet Matters, around $80\%$ of parents in the United Kingdom (UK) use at least one form of parental control, with broadband controls and mobile phone applications (``apps'') being among the most widely adopted solutions~\cite{Wood2023InternetMatters}.

        The prevalence of smartphones and tablets among minors has heightened the importance of parental control apps and made these tools commonplace~\cite{Marciano2022Parental, Mavoa2023SafetyNotSnooping}. 
        Studies show that some parents view monitoring tools not only as helpful instruments but as a necessity for responsible parenting~\cite{Marx2010Beginning,PageJeffery2021Difficult}. This form of ``care surveillance'' has been studied extensively~\cite{Burikova2024Care, Zou2024CrossContextual, Dungey2024BirdsEyeView}. It is reinforced by research indicating that a lack of parental control over a minor’s smartphone escalates the risk of children becoming victims of online harassment~\cite{Schmuck2023Control}, while understanding how children interact with digital media can help reduce family conflict~\cite{Kim2017Tweens}.
        
        While there are many reasons for parents to consider using such apps, concerns have been raised about security and privacy violations in child monitoring products. Feal et al.~\cite{Feal2020AngelOD} discovered that parental control apps lack transparency and do not comply with regulatory requirements. Ali et al.~\cite{Ali2020Betrayed} and Largent~\cite{Largent2017Vulnerability} disclosed several vulnerabilities in such software, including remote code execution, authentication bypass and device malfunction. Whittaker~\cite{Whittaker2018teen} reported on a major data breach leaking thousands of user passwords in a teen phone monitoring app.
        In response to such findings, children's rights advocates, including the 5Rights Foundation, criticised the absence of comprehensive online protection legislation for children, condemning the existence of products and services that pay little regard to the welfare of young people and minors~\cite{5Rights2022Child}. 
        
        Deploying parental control systems unavailable on official platforms, such as the Google Play Store, calls for special consideration. Users can acquire alternative apps by downloading them from third-party websites and installing them on their device after granting the necessary permissions.
        These so-called \textit{sideloaded apps} offer parents additional functionalities such as extensive uninstallation restrictions and hiding their presence through stealthiness and obfuscation~\cite{Kotzias2020UnwantedAppDistribution}.
        Users of such sideloaded services can therefore implement parental controls in scenarios where their child opposes or circumvents monitoring measures~\cite{Wei2022TikTok,Kaufmann2021Secret}.    
        
        Sideloaded apps can introduce additional privacy and security risks that parents might not be aware of. The providers of sideloaded services are not required to follow app store policies, which restrict or prevent the misuse of Android Application Programming Interface (API) capabilities, remote control and similar functionalities~\cite{GooglePlayProtect2023Malware}. 
        Both Apple and Google review apps before allowing them on their stores. Apple uses a team of experts that rejected $27\%$ of apps in 2022, while Google relies on automation and real-time scanning for security~\cite{Morton2024EntryCompetition}.
        Sideloaded apps operate without the safeguards and standards set by such platforms, increasing the risks of data leakage, outdated software and the absence of robust encryption protocols.
        
        Another consideration relates to parents inadvertently installing dual-use apps or stalkerware/spyware\footnote{\ There are different definitions for the terms ``stalkerware'' and ``spyware'', and there is currently no consensus on their key features~\cite{Bermudez2020Mapping}. However, spyware is generally a broader term encompassing various types of software, including adware, keyloggers and rootkits~\cite{Fortinet2023Spyware}. We use ``stalkerware'' to characterise any software falling under the mentioned categories, except when we cite articles explicitly using the term ``spyware''.} marketed as parental control tools on their children's devices~\cite{Mavoa2023SafetyNotSnooping}. Not all parents may fully understand the implications of using sideloaded apps, especially when encountering them through channels such as advertisements or recommendations outside official app stores. While both stalkerware and parental control apps can monitor a person's activities, the key difference lies in a product's aim for secrecy and intent: Stalkerware often conceals its presence and operates without the user's knowledge or consent~\cite{Kaspersky2023Stalkerware}. Indicators of compromise (IOC) for stalkerware (see \autoref{subsec:stalkerwareioc}) have therefore been collected and provided by different sources~\cite{Kaspersky2021Raising, WARNE2024Stalkerware}.

        Given the potential hazards of these technologies and the need for heightened protections for vulnerable groups like children~\cite{5Rights2022Approaches}, our paper seeks to shed light on the sideloaded parental control app ecosystem. Sideloaded apps occupy a unique and underexplored niche in the broader app landscape. Vendors often justify sideloading as a means to navigate challenges posed by official app stores, including strict regulatory requirements. These apps also provide parents with tools that are harder for children to outmanoeuvre and allow vendors to avoid app store fees~\cite{Brady2024AppSideloading, Morton2024EntryCompetition}.
        Despite their prevalence, precise data on the size and impact of this sideloaded app domain remains scarce (see \autoref{subsec:popularity}).
        
        Our focus is specifically on Android-based systems, which held approximately $70\%$ market share in the first quarter of 2024~\cite{Statista2024MobileOS} and where sideloading is a common and relatively straightforward process. By contrast, sideloading is far less prevalent on iOS devices, where app distribution is tightly controlled, and sideloading requires rooting~\cite{Greene2018Platform}. Yet, this dynamic may shift as the European Union's (EU) Digital Markets Act (DMA)\footnote{\ The DMA restricts the influence of major digital platforms acting as gatekeepers by mandating that companies like Apple permit software from outside their app stores and submit annual reports on DMA compliance~\cite{DMA2024}.} is increasingly enforced~\cite{Apple2024DMA}. This regulatory backdrop, coupled with the distinct risks associated with sideloaded apps, highlights the urgency of examining this ecosystem in depth.
        The driving research questions for our study, therefore, were:
        \begin{enumerate}
            \item[Q1:] How do sideloaded parental control apps compare to Google Play Store parental control apps in terms of permissions, trackers, functionalities and obfuscation techniques?
            \item[Q2:] How clearly and transparently do sideloaded parental control apps communicate their privacy and data collection practices to users within their privacy policies?
            \item[Q3:] How do sideloaded parental control apps handle encryption and sensitive data transmission?
            \item[Q4:] How many sideloaded parental control apps match common stalkerware indicators of compromise (IOC)?
        \end{enumerate}


\section{Background}
        Parental control apps serve as an element of guardianship, frequently acting as a protection mechanism for parents to ensure a safe and secure online environment for their children.
        Indeed, parents currently face various challenges in their duty of care.
        Minors who engage with internet-enabled services can be subjected to harmful behaviour and may end up being confronted with cyberbullying, online grooming, inappropriate content and various forms of scams~\cite{Boyd2023Connected,Throuvala2021Policy,PageJeffery2021Difficult}. Teens who have encountered such issues are also more likely to be closely supervised by their parents~\cite{Ghosh2018Matter} and contextualise these privacy-invasive behaviours differently when they occur via electronic means~\cite{Laird2024Perceptions}. 
        
        \subsection{Dataveillance}
        A downside of children being monitored by and engaging with digital technologies is that they have become subject to many surveillance systems, emphasising the urgency of addressing children's rights and holding industry actors responsible.
        Technologies spanning from mobile games and wearables to educational software generate extensive data about children, capitalising on and profiting from their personal information~\cite{Lupton2017Datafied}.
        While children may not always fully understand practices of datafication,\footnote{\ ``Datafication'' here includes the recording, tracking, aggregating, analysing and exploiting of children's online actions for purposes such as monetisation~\cite{Wang2015Reevaluating}.} they care about its various aspects and want to be able to stop their data from being used in ways unknown to them~\cite{Wang2022Assumptions}.
        Children’s rights instruments that address the risks and harms of data surveillance remain an area for improvement~\cite{Lupton2017Datafied}, with UNICEF arguing for children’s data protection responsibility also falling onto tech vendors~\cite{UNICEF2020Rights}.

        \subsection{Misuse}
        Besides the risks associated with parental control apps processing children's data, concerns arise about their implemented monitoring capabilities that make them susceptible to misuse.
        Child monitoring apps often offer features like location tracking, message monitoring and call log access – functionalities that could potentially be exploited by malicious actors~\cite{Khoo2019Fear}.
        This dynamic raises the critical issue of parental control apps serving as a dual-use technology.
        Chatterjee et al.~\cite{Chatterjee2018spyware} define dual-use apps as software designed for a legitimate use case that can be misused by an abuser, such as a perpetrator of intimate partner violence (e.\,g.\ because their capabilities permit another person remote access to a device's data without the device user's knowledge). 
        Past literature has established that some parental control apps, as well as Bluetooth Trackers, can and have been repurposed~\cite{Chatterjee2018spyware,Freed2018Stalkers,Molnar2019Consumer, Heinrich2024UnstalkMe}, especially if not enough precautions are put in place~\cite{Khoo2019Fear}. 

        \subsection{Marketing Shift}
       Some stalkerware products have been revealed to advertise themselves or particular functionalities as parental control~\cite{Molnar2019Consumer,Citron2015Spying,Chatterjee2018spyware}. This is problematic, as these products not only mislead users but also typically invest minimal effort in securing the sensitive data they collect~\cite{Liu2023No, WARNE2024Stalkerware, Gibson2022Monetization}. Recent data breaches and court rulings confirm this~\cite{TechRadar2024Offline}. For instance, the hacking of two stalkerware firms in 2018 exposed private text messages containing conversations between children and their parents or spouses. One text read ``You cheated\dots.smh\dots'', while others appeared to be written by children about issues at school~\cite{Cox2018Hacker}. Furthermore, the United States of America (US) Department of Justice's 2014 criminal indictment against the makers of the spyware app StealthGenie appears to have prompted some vendors to shift their marketing focus. Rather than targeting ``cheating partners'', they now emphasise the more ``legitimate'' monitoring of minors or employees~\cite{Liu2023No}.

\section{Related Work}
    This section summarises the literature on in-store parental control and children's apps. Since there is limited research on the sideloaded parental control market, we also include studies that focus on stalkerware apps, as many of these are marketed as parental control tools.
    Additionally, we look at past publications on recommendations for parental control software developers.

    \subsection{Research on In-Store Parental Control Apps}
        In-store childcare and children's apps showcased a lack of transparency and compliance with regulatory requirements. Feal et al.~\cite{Feal2020AngelOD} studied the Android parental control app ecosystem and discovered that $34\%$ of the tested apps sent personal information without appropriate consent and $72\%$ shared data with third parties without stating this in their privacy policies. They raise the question of whether protecting children justifies the risks regarding the collection and processing of their data.
        Oko\-yo\-mon et al.~\cite{Okoyomon2019Contradictions} substantiated these findings by observing that $10\%$ of over 68,000 apps on the Google Play Store's Designed for Families program share data, including personal identifiers, with third parties without declaring this in their privacy policies. By conducting a privacy policy analysis, they also found that $9.1\%$ did not acknowledge that they targeted kids or did not even mention them in their privacy policy.
        Additionally, interviews have shown that many developers lack a good understanding of their own apps' data collection behaviour due to the use of third-party software development kits,\footnote{\ A software development kit (SDK) is a set of tools and resources that enables developers to create apps for a specific platform or framework.} often leading to unintentional violations of privacy regulations~\cite{Alomar2022Darnedest}.
        
        Many children-focused apps use tracking libraries and require a large number of permissions, including ones classified as dangerous by the Android protection levels. For instance, Sun et al.~\cite{Sun2023Measuring} concluded that $81.25\%$ of family apps use trackers, while Gruber et al.~\cite{Gruber2022Diaper} found that $45\%$ use more than two trackers. 
        Concerning permissions, Gruber et al.~\cite{Gruber2022Diaper} calculated a mean of 19~potentially dangerous permission per app. Sun et al.~\cite{Sun2023Measuring} discovered that $4.47\%$ of children’s apps request location permissions, despite the platforms prohibiting the collection of children's location data.

        Researchers have brought to light a myriad of vulnerabilities that undermine children's online and offline safety. 
        Ali et al.~\cite{Ali2020Betrayed} exposed pervasive security and privacy issues, including improper access control, the possibility of online password brute-forcing, vulnerable backends and lack of encryption.
        Blancaflor et al.~\cite{Blancaflor2021OWASP} also detected apps sending unencrypted data and further identified 18~unique vulnerabilities in three popular apps based on Open Worldwide Application Security Project (OWASP) security requirements.
        
    \subsection{Research on Dual-Use Apps and Stalkerware}
    \label{subsec:research-stalkerware}
        Prior research has extensively documented the entrenched issue of dual-use and stalkerware apps, particularly within the realms of parental control and domestic abuse. Almansoori et al.~\cite{Almansoori2022Survey} showcased the prevalence of dual-use apps exploited in intimate partner surveillance on the Google Play Store, identifying 3,988 out of 51,868~apps as dual-use, with many categorised as parental control. They noted that many more dual-use apps may be distributed outside official platforms, calling for future research into the prevalence of sideloaded versions. 
        Chatterjee et al.~\cite{Chatterjee2018spyware} explored apps used in intimate partner violence cases, including 23~sideloaded apps unavailable on official stores, many of which belonged to the child monitoring and subordinate tracking categories. 
        
        Investigations have also scrutinised the promotion and monetisation tactics employed by spyware vendors. Molnar and Harkin~\cite{Molnar2019Consumer} found that spyware is frequently marketed for both malicious and ostensibly ``credible'' purposes, such as parental control -- though the monitoring capabilities often exceed appropriate levels. Their content analysis revealed that multiple apps advertising child monitoring also promoted spying on intimate partners. Complementing this, Gibson et al.~\cite{Gibson2022Monetization} analysed over 6,400 spyware apps, showcasing not only their diverse revenue generation strategies but also the broader ecosystem of payment processors that sustains this surveillance industry. 
        Researchers have further exposed security flaws and data protection failures within these types of apps. Technical analyses of stalkerware by Liu et al.~\cite{Liu2023No} and Mannan and Youssef~\cite{Mannan2023IPA} included apps advertised as parental control, providing insights into the abuse of Android APIs, functionalities and security shortcomings. 
        
        Our present work builds on these prior studies, examining how the parental control market overlaps with stalkerware.
        Six of the apps reviewed by Liu et al.~\cite{Liu2023No} are also part of this paper, as well as eleven of the apps analysed by Mannan and Youssef~\cite{Mannan2023IPA}. 
        By expanding on their research and incorporating an examination of privacy policies and IOCs, our research offers additional insights and extends their findings.
        
    \subsection{Recommendations and Best Practices}
        In the face of the unveiled issues in children's and parental control software, regulators created recommendations for software developers handling children's and other vulnerable communities' data~\cite{Brown2024IoT}.
        The eSafety Commissioner, an independent Australian regulator~\cite{eSafetyCommissioner2019Principles}, formulated safety-by-design principles. These emphasise the responsibility of service providers to reduce their users' exposure to harm. They further demand user empowerment and autonomy to support safe online interactions and advocate for transparency and accountability in the tech sector. Other resources propose similar measures, such as the child-rights-by-design principles created by the aforementioned 5Rights Foundation~\cite{Livingstone2023Rights}.
        
        Literature has also provided guidance and best practices for parental control.
        Gnanasekaran and De Moor~\cite{Gnanasekaran2023Recommendations} conducted a literature review which yielded nine recommendations, split into the categories of usability, security and privacy. The security recommendations include increasing awareness amongst children and parents and following security standards and procedures. The privacy recommendations suggest monitoring children's activity in non-intrusive ways, making it clear what parents consent to in terms of data sharing and granting parents access to the children's data to delete or restrict data collection by an app.

    \subsection{Research Gap}
        While the Background section highlighted that there are publications on in-store parental control apps and stalkerware, including dual-use parental control, little is known about the sideloaded parental control industry. 
        Prior work on parental control apps primarily focused on software available through official app stores~\cite{Feal2020AngelOD, Okoyomon2019Contradictions}, excluding sideloaded software. This lack of research implies the existence of an evidence gap in the understanding of the risks, privacy concerns and functionalities associated with the sideloaded parental control ecosystem.
        Considering the issues uncovered by prior literature, it can be assumed that the situation for sideloaded apps is likely similar, if not worse. 
        Sideloading presents increased risks due to issues unique to sideloaded apps, including the dearth of independent oversight by an official app store, the lack of user ratings and reviews to inform users and the inability to provide regular, automatic updates.
        Besides, the absence of robust studies, which scholars like Almansoori et al.~\cite{Almansoori2022Survey} called for, may hinder the development of informed policies and guidelines for the responsible design and the ethical use of such apps moving forward.

\section{Methodology}
        We selected a comparable sample size of sideloaded and Google Play Store parental control apps. We then assessed these apps against stalkerware indicators of compromise (IOC) provided by TinyCheck's~\cite{TinyCheck2024} GitHub repository. A static analysis was conducted on both sideloaded and in-store apps, followed by a comparative analysis of their features. We further collected and analysed their network traffic data and privacy policies and compared the data-sharing practices outlined in the policies with the actual network traffic observed.
        The data collection encompassed the period from May to November 2024.
        
    \subsection{Selection Process of Sideloaded Apps}
        We conducted a Google web search using a list of 15 phrases (see \autoref{appendix:selectionterms}) that parents might use when seeking a legitimate parental control solution. To exclude results from the Google Play Store, we added the term ``-site .google.com''. We manually analysed the first 60 results of each search query, as Google automatically stops loading websites after this point and the results tend to be less relevant. Any tools mentioned across a cumulative total of 900 results were documented to obtain an extensive list of all currently relevant software in the parental control software market, resulting in a total of 132 tools. We further narrowed this set of software down to only include Android apps that are not available on the Google Play Store. After subtracting tools that were no longer accessible, we were left with 44 Android apps. These were further inspected to exclude apps that redirect to or were purchased by other parental control services already on the list. Excluded were Refog (part of Aura, available on Google Play Store), Circle (redirects to FlexiSpy), MobileSpy (redirects to mSpy) and Spynger\footnote{\ On their website, Spynger advertises itself as the ``best cheating app'' and promotes spying on your partner if there are suspicions of infidelity.} (acquired by mSpy, as confirmed by their support staff).
        Subsequently, the list was sorted according to the Tranco rankings of each app's official website~\cite{LePochat2019} (see \autoref{subsec:popularity}), with 20 websites ranking among the top 1 million most visited and popular sites on the internet. Due to their lower visibility and relevance, apps without a Tranco score (i.\,e.\ not among the first 1 million websites) were disregarded. 
        Eleven apps that did not offer a free version or trial period were purchased, with prices ranging from €23.28 to €59.99. Paying for the apps was necessary to ensure a representative analysis of the sideloaded market without omitting popular options. While this included transactions with companies that could be considered as offering stalkerware (see \autoref{subsec:stalkerwareioc}), it was necessary to gain insights into the potential risks associated with such software and to advance the current knowledge on the subject.
        The resulting 20 apps analysed in this paper were Bark, mSpy, Spapp Monitoring, Hoverwatch, AnyControl, uMobix, TheOneSpy, Kidlogger, FlexiSpy, XNSPY, TiSpy, SPYX, WebWatcher, MoniMaster, Cocospy, iKeyMonitor, SPY24, Chyldmonitor, EvaSpy and Kidstracker (listed in order of descending Tranco ranking).

    \subsection{Selection Process of Google Play Store Apps}
        To select Google Play Store apps, we employed the same search terms used for sideloaded apps (see \autoref{appendix:selectionterms}). The results were manually filtered to identify apps intended for child devices, replacing parent versions with their child counterparts. During this process we encountered two cases where the parent app was available on the Google Play Store while the child app could only be downloaded from the vendor's website: AirDroid Parental Control and FlashGet Kids: Parental Control. Prior to selecting the apps with the highest download numbers, we excluded apps tailored to one specific service, such as the Nintendo Switch Parental Controls or YouTube Kids. Two more apps were excluded because they offered significantly fewer features, limited to GPS tracking and few additional functionalities, making them unsuitable for direct comparison with both sideloaded and in-store parental control apps.
        We picked a sample size of 20 apps, corresponding to the number of sideloaded apps: Life360, iSharing: GPS Location Tracker, Pingo, Kids Place Parental Control, Kaspersky SafeKids with GPS, Alli360, Parental Control – Kidslox, ST Kids App, Spy Phone Labs Phone Tracker, KidControl Family GPS locator, Kids App Qustodio, MMGuardian Child Phone App, Parental Control Kroha, Microsoft Family Safety, ESET Parental Control, MobileFence, Tigrow!, FamiSafe Kids, FamilyTime Jr.\ and Kidsy (listed in descending order of approximate download numbers from the Google Play Store).

    \subsection{Popularity}
    \label{subsec:popularity}
        To estimate the popularity of sideloaded parental control, we utilise the Tranco rankings taken on 2~June 2024 because app usage numbers reported on websites are considered unreliable as they tend to exaggerate their customer base~\cite{Molnar2019Consumer}. Of the 20 apps, the most popular one is Bark with a score of 10,840. To put this number into perspective, \url{samsungmobile.com} (scoring 10,856) and \url{nintendo.de} (scoring 10,835) rank similarly. Four apps are ranked among the top 100,000 websites, nine among the top 300,000 and 16 among the top 500,000.
        Another metric that can be used to estimate the prevalence of these tools is data breaches: security researchers found over 60,000 users registered with XNSpy~\cite{Whittaker2022Xnspy}, and another breach unveiled 130,000 accounts on Retina-X and FlexiSpy~\cite{FranceschiBicchierai2017Inside}. A recent incident in May 2024 exposed the databases of mSpy, leaking almost 2.4~million unique emails~\cite{Apostol2024mSpy}.
        
    \subsection{Setup}  
        The setup consisted of an emulated Android device with an API level of 31 which corresponds to Android 12.0. This API level was chosen due to being widely supported by developers, increasing the likelihood of the app features working on the emulator.\footnote{\ The AnyControl app had to be installed on a physical device (using Android 12.0) because it was built for ARM architecture, which is incompatible with an x86\_64 emulator.}
        For each sideloaded app a new emulated device was set up and the parental control app was subsequently installed using the APK file (i.\,e.\ Android package kits, used to distribute apps on the Android operating system) downloaded from the vendor's website. In-store parental control apps were tested on a physical device running Android 9.0. None of the studied apps required a rooted device. The vendors of sideloaded apps typically provided a dashboard displaying the monitored data on their website, while in-store vendors were more likely to offer a second app or mode for parents.
        Charles Proxy~\cite{CharlesProxy2024} was used to capture and analyse network traffic, including HTTPS traffic. To decrypt the traffic, Charles Proxy certificates were installed on the emulators. For apps using certificate pinning\footnote{\ A security technique where the app is configured to only trust a specific certificate, thus preventing the capture of unencrypted traffic with the help of unauthorised certificates.}, we used Frida~\cite{Frida2024}, a toolkit that injects custom scripts into running apps to bypass certificate pinning.
        All apps were downloaded and tested in Austria, consequently requiring the vendors to adhere to the General Data Protection Regulation (GDPR), which applies to all mobile apps that collect and process the personal data of EU citizens (regardless of whether the processing occurs in the EU)~\cite{EuropeanParliament2018GDPR3}. The emulator's GPS was enabled and set to a location in Austria.
        
        The apps were installed in line with the installation guides provided by the parental control vendors.
        The features of each app were tested by manually interacting with the phone for approximately thirty minutes. In this period, user behaviour was mimicked by engaging with different installed apps, browsing websites, downloading content, taking pictures, receiving and making calls, receiving and writing SMS messages and so on.

    \subsubsection{Obtaining Google Play Store APKs}
        To conduct a static analysis, we first required the APK files of the Google Play Store apps. We employed an Android emulator using a Google Play image, which allowed us to install the selected apps on the emulator directly. Using the Android Debug Bridge (adb)~\cite{ADB2024}, we then determined the app package and its path and extract it from the emulator. In the case of split APKs, which are multiple APK files making up one app to optimise file size and delivery for different device configurations, we merged the split files using the open-source tool APKEditor~\cite{APKEditor2024}.

    \subsection{Privacy Policies}
        The presence of a privacy policy on the vendor's website, in the Google Play Store or in the app was evaluated to ensure that both parties involved, such as parents and children, receive clear information regarding the app's privacy practices. The policies were analysed in a word-by-word textual examination.

        We extracted the following information from the policies: length, Flesch reading ease score~\cite{Flesch1948Readability}, last update, scope, type of collected data, sharing with third parties including mentioned third parties and reasons for sharing, user rights and contact info. This analysis was performed manually.
        We further categorised the policy scope into five groups to assess its relevance to the user. This is a slight modification of the categorisation used by Sunyaev et al.~\cite{Sunyaev2015Availability}: 
        \begin{enumerate}
            \item Website only: Policy covers only the website.
            \item App only: Policy covers only the app under test.
            \item Multiple apps: Policy covers multiple apps by the same vendor, including those under test.
            \item All company-based services: Policy covers all services the vendor offers.
            \item No relation: No relation was found between the policy and the app under test.
        \end{enumerate}

        The Flesch reading ease score assesses the ease of understanding a written text based on sentence length and the number of syllables per word.
        It ranges from 0 to 100; the higher the score, the easier the text is to understand.
        The online tool Flesch Kincaid Calculator~\cite{FleschReadingEase2024} was used for the calculation.

        \subsection{Static Analysis}
        To conduct a static analysis, we reviewed the APK files with Exodus~\cite{Exodus2024} and MobSF~\cite{MobSF2024}, which provided output for deeper examination. 
        Security researchers frequently employ MobSF to conduct security analyses on mobile apps~\cite{Papageorgiou2018Analysis} as it can identify a wide range of Android security issues~\cite{Ranganath2020Tools}.
        Exodus has also been used by cybersecurity companies~\cite{Kaspersky2019Exodus} and researchers~\cite{Laperdrix2022Exodus}.
        We used both tools to conduct a static analysis of the parental control apps without executing them. This involves automated analysis of the manifest files, source code and assets to identify specifications and security vulnerabilities. We compared the generated lists of permissions, app libraries, trackers and general technical information wherever applicable. We further split the permissions into potentially dangerous permissions and other categories, based on the protection levels specified by Android~\cite{AndroidForDevelopers2023Permissions}.

\section{Results}
    The next section presents our analysis of 20 sideloaded parental control apps and 20 Google Play Store parental control apps. We first outline the results of our app analysis, followed by the examination of their privacy policies and network traffic. Last, we spotlight our findings regarding stalkerware IOC. 


        \subsection{Static Analysis}
        All APK files were subjected to static analysis using both Exodus and MobSF. We compared the results from these tools and resolved discrepancies. This included cases where the number of trackers identified differed between the two tools or where MobSF detected additional permissions that Exodus did not.
        The key findings from this analysis are presented  in two separate tables. \autoref{table:instore} covers the apps available on the Google Play Store and \autoref{table:sideloaded} focuses on sideloaded apps.



        \subsubsection{Trackers}
        Our analysis revealed that Google Play Store apps generally have more trackers than sideloaded apps. On average, sideloaded apps had 1.8 trackers, with a range from zero to five, whereas in-store apps averaged 4.2 trackers. iSharing: GPS Location Tracker had the most trackers, totalling nine.
        We identified 15 different trackers among the sideloaded apps. The most frequently used were Google CrashLytics, present in nine apps, and Google Firebase Analytics, which appeared in eleven apps.
        The Google Play Store apps contained 22 unique trackers. Google CrashLytics and Google Firebase Analytics remained the most popular ones, with 18 and 16 occurrences respectively. A more detailed breakdown can be found in \autoref{table:tracker}.
        
        \begin{table}[ht]
        \centering
        \caption{Static Analysis of Google Play Store Apps}
        \label{table:instore}
        \resizebox{0.4\textwidth}{!}{
        \small
        \begin{tabular}{l c c c c c c}
        \rotatebox{90}{\textbf{App Name}} & 
        \rotatebox{90}{\textbf{\shortstack{Downloads}}} & 
        \rotatebox{90}{\textbf{\shortstack{Permissions}}} & 
        \rotatebox{90}{\textbf{\shortstack{Dangerous \\ Permissions}}} & 
        \rotatebox{90}{\textbf{Tracker}} & 
        \rotatebox{90}{\textbf{Obfuscation}} & 
        \rotatebox{90}{\textbf{IOC Match}} \\
        \hline \hline
        Alli360 & 1M+ & 34 & 10 & 4 & N & \\ \hline
        ESET Parental Control & 1M+ & 31 & 10 & 1 & N & \\ \hline
        FamilyTime Jr & 100k+ & 37 & 13 & 3 & N & \\ \hline
        FamiSafe Kids & 500k+ & 38 & 15 & 8 & N & \\ \hline
        iSharing & 10M+ & 35 & 16 & 9 & N & \\ \hline
        Kaspersky SafeKids & 1M+ & 48 & 8 & 6 & N & \\ \hline
        KidControl & 1M+ & 25 & 6 & 4 & N & \\ \hline
        Kids Place & 5M+ & 37 & 10 & 4 & N & \\ \hline
        Kidslox & 1M+ & 42 & 16 & 7 & N & \\ \hline
        Kidsy & 100K+ & 33 & 12 & 3 & N & \\ \hline
        Kroha & 1M+ & 39 & 12 & 3 & N & \\ \hline
        Life360 & 100M+ & 42 & 18 & 5 & N & \scalebox{0.8}{\ding{108}} \\ \hline
        Microsoft Family Safety & 1M+ & 22 & 9 & 3 & N & \\ \hline
        MMGuardian & 1M+ & 45 & 17 & 7 & N & \\ \hline
        MobileFence & 1M+ & 58 & 19 & 2 & N & \\ \hline
        Pingo & 10M+ & 35 & 12 & 3 & N & \scalebox{0.8}{\ding{108}} \\ \hline
        Qustodio & 1M+ & 27 & 5 & 3 & N & \\ \hline
        \text{\small Spy Phone Labs Phone tracker} & 1M+ & 15 & 5 & 0 & N & \scalebox{0.8}{\ding{108}} \\ \hline
        ST Kids App & 1M+ & 27 & 11 & 7 & N & \\ \hline
        Tigrow & 500k+ & 27 & 12 & 1 & N & \\ \hline
        \hline
        \multicolumn{7}{c}{\small \textbf{Abbreviations:} N = No, IN = Icon and Name, NO = Name Only} \\
        \end{tabular}
        }
        \end{table}

            \subsubsection{Permissions}
            Compared to parental control apps available on the Google Play Store, sideloaded parental control apps request significantly more permissions. On average, sideloaded apps requested 44.4 permissions, whereas in-store apps averaged 34.9 permissions.
            The range of requested permissions for sideloaded apps varied from 22 to 70, with mSpy requiring the most. For Google Play Store apps, the range was from 15 to 58 permissions, with Mobile Fence requesting the most.
    
            Regarding permissions classified as dangerous by Android, sideloaded apps similarly request more, averaging 21 dangerous permissions compared to 11.8 for in-store apps. Among Google Play Store apps, MobileFence – Parental Control requested the most dangerous permissions with 19, while the sideloaded app mSpy requested the most with 31.

            We took a closer look at the permissions required by some of the apps. We identified several permissions that appear unnecessary for parental control or do not align with the app's features, including Bluetooth access, user credentials access, superuser permissions, flashlight access and dictionary modifications. 

            \begin{table}[ht]
            \centering
            \caption{Static Analysis of Sideloaded Apps}
            \label{table:sideloaded}
            \resizebox{0.4\textwidth}{!}{
            \small
            \begin{tabular}{l c c c c c c}
            \rotatebox{90}{\textbf{App Name}} & 
            \rotatebox{90}{\textbf{\shortstack{Tranco Score}}} & 
            \rotatebox{90}{\textbf{\shortstack{Permissions}}} & 
            \rotatebox{90}{\textbf{\shortstack{Dangerous \\ Permissions}}} & 
            \rotatebox{90}{\textbf{Tracker}} & 
            \rotatebox{90}{\textbf{Obfuscation}} & 
            \rotatebox{90}{\textbf{IOC Match}} \\
            \hline\hline
            AnyControl & 135,169 & 53 & 24 & 2 & IN & \\ \hline
            Bark Premium & 10,840 & 44 & 14 & 4 & N & \\ \hline
            Chyldmonitor & 728,971 & 38 & 21 & 1 & IN & \\ \hline
            Cocospy & 461,897 & 38 & 21 & 1 & IN & \scalebox{0.75}{\ding{108}} \\ \hline
            EvaSpy & 775,227 & 44 & 17 & 0 & IN & \\ \hline
            FlexiSpy & 264,503 & 47 & 24 & 1 & IN & \scalebox{0.75}{\ding{108}} \\ \hline
            Hoverwatch & 84,407 & 41 & 25 & 0 & IN & \scalebox{0.75}{\ding{108}} \\ \hline
            iKeyMonitor & 498,000 & 57 & 25 & 3 & IN & \scalebox{0.75}{\ding{108}} \\ \hline
            Kidlogger & 264,485 & 22 & 12 & 0 & N & \\ \hline
            Kidstracker & 832,453 & 30 & 15 & 1 & IN & \\ \hline
            MoniMaster & 446,047 & 34 & 19 & 2 & IN & \\ \hline
            mSpy & 54,619 & 70 & 31 & 4 & NO & \scalebox{0.75}{\ding{108}} \\ \hline
            Spapp Monitoring & 80,573 & 45 & 24 & 0 & IN & \scalebox{0.75}{\ding{108}} \\ \hline
            SPY24 & 501,038 & 53 & 26 & 1 & IN & \\ \hline
            SPYX & 419,471 & 52 & 23 & 0 & IN & \\ \hline
            TheOneSpy & 260,677 & 40 & 18 & 2 & IN & \\ \hline
            TiSpy & 404,848 & 44 & 17 & 2 & IN & \scalebox{0.75}{\ding{108}} \\ \hline
            uMobix & 256,352 & 56 & 28 & 4 & IN & \\ \hline
            WebWatcher & 424,303 & 38 & 16 & 2 & IN & \\ \hline
            XNSPY & 365,926 & 41 & 18 & 5 & NO & \scalebox{0.75}{\ding{108}} \\ \hline
            \hline
            \multicolumn{7}{c}{\small \textbf{Abbreviations:} N = No, IN = Icon and Name, NO = Name Only} \\
            \end{tabular}
            }
            \end{table}
            
            We further analysed three special permissions – Accessibility, Device Administrator and Draw Over Other Apps – that protect access to particularly powerful actions.
            
            Every single sideloaded app leveraged the Accessibility permission, intended to help users with disabilities, for example by creating screen readers or improving text readability. In parental control apps, this feature was used to view content displayed on the phone screen and control device interactions. Among the in-store apps, 14 used the Accessibility permission, while 6 did not.
            
            The Device Administrator permission was used by 16 sideloaded apps and 15 in-store apps. This permission grants the ability to change passwords and lock the device but was mainly used to hinder the parental control app's uninstallation in the apps analysed by us. Device Administrator apps cannot be uninstalled like normal apps; the permission must be revoked before the app can be uninstalled.
            
            The last special permission, Draw Over Other Apps, was used by 13 sideloaded apps and 14 in-store apps. This permission can be used to display content on top of other apps. We observed this permission being mainly used to further hinder the parental control app's uninstallation, for example, by automatically closing the settings app when an uninstallation attempt is made.
        
        \subsection{Functionalities}
        We documented the functionalities of each app during testing, including features not available in our current subscription model based on the information given on the websites or on the Google Play Store. 
        The functionalities provided by the analysed parental control apps can be divided into the following categories:
            \begin{enumerate}
                \item \textbf{Monitoring capabilities:} This encompasses basic monitoring functionalities provided by most parental control apps, including reading SMS messages, call history and browser history or location tracking.
                \item \textbf{Social media and messengers:} The majority of analysed apps offered the monitoring of popular social media apps and messengers. The most common were Facebook, Instagram, Snapchat and WhatsApp. 
                \item \textbf{Remote control and access:} Remote control could be exerted via the monitoring dashboard of the parental control service or with SMS codes. Capabilities include making the device ring, activating the cameras or microphone and taking screenshots. 
                \item \textbf{Specialised parental control:} This category contains functionalities that are typically only useful for parental control and have limited use for, for example, intimate partner surveillance, such as time limits and content filtering.
            \end{enumerate}
            Expanding on the last category, we defined a list of functionalities that fall under specialised parental control: app time limits, setting schedules, content filtering, SOS/panic buttons, geofencing, keyword detection,\footnote{\ Apps such as Bark monitor keywords and send a report if a term appears in conversational texts or other activity. This can be used to detect cyberbullying, access to inappropriate content etc.} homework/task management, blocking apps and spyware detection.
            Given past research showing that stalkerware has presented itself as sideloaded parental control~\cite{Molnar2019Consumer,Citron2015Spying,Chatterjee2018spyware}, we determine whether the analysed apps include specialised parental control functionalities and use this to indicate whether parental control was an objective for the app developers. 
            
            Five sideloaded apps did not include any specialised parental control functionalities: Cocospy, FlexiSpy, Hoverwatch, Kidstracker and Webwatcher.
            In comparison, none of the Google Play Store apps lacked specialised parental control features.

            We also observed that Google Play Store apps offered fewer functionalities than sideloaded apps. Monitoring of other installed social media apps, which is heavily advertised by all sideloaded apps, is only mentioned by four in-store apps. Other features, such as keylogging, remote camera access and screen recordings were only offered by sideloaded apps. 

            Another concerning finding is that 12 of the sideloaded parental control apps promoted the monitoring of dating apps such as Tinder and Hinge, which could be used to monitor adults, as both apps have an age requirement of 18 years.

            \subsection{Obfuscation}
            A frequently quoted differentiator between parental control apps and stalkerware is whether the app hides its activity from the device owner~\cite{Kaspersky2023Stalkerware}. Therefore, we examine which apps obfuscate their names and icons to appear inconspicuous or to masquerade as a different service.

\begin{figure}[ht]
            \centering
            \includegraphics[width=0.7\linewidth]{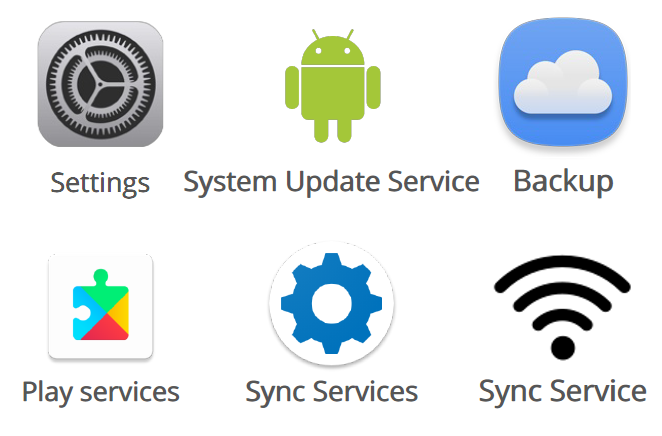}
            \caption{Obfuscation of sideloaded apps (upper left to lower right): AnyControl, MoniMaster, Spapp Monitoring, uMobix, FlexiSpy and Hoverwatch}
            \Description[Image containing examples of obfuscated app icons]{The images resemble icons for settings, the Android logo, a cloud symbol and a Wi-Fi symbol}
            \label{img:obfuscation}
            \end{figure}

            None of the analysed in-store parental control apps used obfuscation -- which is unsurprising considering it is prohibited by the Google Play Store policies~\cite{Google2024Malware}. 
            Contrarily, 17 out of 20 sideloaded apps obfuscated their name; the only exceptions were Bark, Kidlogger and WebWatcher. Common obfuscated names include generic or system-related terms like ``Settings'', ``System Service'', ``WiFi Service'' and ``Sync Services''. These names are designed to blend in with legitimate system functions, giving the impression that they are a standard component of the device's software.
            Regarding the sideloaded app icons, 17 apps used an innocuous image, such as the ones shown in \autoref{img:obfuscation}, i.\,e.\ all except for Bark, Kidlogger and mSpy\footnote{\ mSpy provided users with the option to obfuscate the app, but it was not enabled by default.}. Oftentimes icons resembling the settings icon, the Android logo or a WiFi symbol were used. 

        \subsection{Google Play Protect Services}
        Google Play Protect is a security feature on Android devices that scans and safeguards against potentially harmful apps and malware from the Google Play Store and other sources. Disabling Google Play Protect leaves the device vulnerable to malware and viruses, which is not ideal, especially for children's phones. However, 17 out of 20 sideloaded apps instruct the user to disable the feature, as otherwise the parental control app might be flagged as malicious and disabled by Google Play Protect.
        We tested how many of the sideloaded apps would be detected by Google Play Protect. In total, 13 apps were detected by Google Play Protect version 42.1.27-31, whereas seven were not considered to be harmful: Bark, EvaSpy, FlexiSpy, Spapp Monitoring, SPYX, TheOneSpy and TiSpy. 

        \subsection{Privacy Policies}
        Our findings reveal significant differences between in-store and sideloaded apps regarding the availability, applicability and completeness of their privacy policies, raising concerns about transparency and user privacy. \autoref{table:policies} offers an overview of key results.
      
        \subsubsection{Availability and Applicability}
        While all 20 sideloaded app vendors included a privacy policy on their website, just eight linked to or displayed a privacy policy within the app and a further eight apps did not seek consent for their privacy policies at all.
        Another issue we identified was that even if consent was sought, several sideloaded apps did not include the app in the scope of their privacy policy. 
        Ten privacy policies applied only to the vendor's websites, leaving the app without a relevant privacy policy. Since the content of these policies was not applicable to our analysis, they were excluded. Only policies for AnyControl, Bark, Hoverwatch, Kidstracker, MoniMaster, Spapp Monitoring, SPY24, uMobix, WebWatcher and XNSPY remained.

        Concerning in-store apps, 19 out of 20 provided an applicable privacy policy on the Google Play Store, and 16 of these included a direct link to the policy within the app itself. Only KidControl failed to provide a privacy policy. Although the app remained available in the store, it was non-functional, with user reviews indicating that it had been broken since around a month prior.

        \subsubsection{Readability}
        The readability of the policies ranged from fairly difficult to very difficult, with an average score of 32.8 on the Flesch–Kincaid readability test for sideloaded apps, corresponding to the reading level of college students or college graduates. 
        The policy of uMobix stands out as an outlier, being the only one to achieve a score above 50.
        The average length of the policies for sideloaded apps was around 17,330 characters. mSpy and SPYX were significantly longer, with around 44,700 and 43,800 characters. Comparatively short are the policies for AnyControl and Kidlogger, with both being less than 5,000 characters in length.

        In-store app policies had comparable readability, with an average Flesch-Kincaid score of 31.1, but were longer, averaging 23,460 characters. The length of in-store policies also exhibited variation, ranging from 3,720 characters for MobileFence – Parental Control to 57,580 characters for Life360.
        
        \subsubsection{Contact Information}
        In-store apps were more likely to specify data protection officer (DPO) contact information, with 11 doing so compared to just one sideloaded app. Excluding those without privacy policies, one app from each group lacked any contact details, while the rest provided only general information.
        The GDPR requires companies that process sensitive data on a large scale or involve large-scale, regular and systematic monitoring of individuals to appoint a DPO~\cite{EuropeanCommissionGuidelines2017DPO}. 
        This applies to the analysed parental control vendors, raising questions of compliance with the GDPR.
        
        \subsubsection{User Rights}
        User rights were mentioned by 15 in-store apps but only by 6 sideloaded apps. This disparity implies that sideloaded apps are more likely to lack full adherence to regulatory requirements.
        According to the GDPR, the types of data collected, the legal basis for processing, the sharing of information and user rights must be communicated~\cite{EuropeanParliament2018GDPR13}. 

        \subsubsection{Data Sharing with Third Parties}
        All ten sideloaded apps and 17 in-store apps stated in their policies that they share data with third parties, though it was often unclear whether personal data was included. Only one sideloaded app specified a third-party recipient, compared to 12 in-store apps that identified the services to which they provided data. Both groups cited similar reasons for data sharing, most commonly compliance with legal and regulatory requirements, service provision and marketing. However, in-store apps demonstrated significantly greater transparency on this topic, while the explanations given by sideloaded apps often lacked specific details, making it difficult to assess the extent of data-sharing practices.
        
        \subsection{Network Traffic}
        \label{subseC:traffic}
        The network traffic captured during the testing was analysed to determine the usage of certificate pinning, encryption and URLs.
        Nine sideloaded apps employed some form of certificate pinning, compared to seventeen in-store parental control apps.
        The higher adoption of certificate pinning in in-store apps is expected, given that sideloaded apps have been found to neglect security standards~\cite{Zimperium2024Sideloading, Apple2021Ecosystem}. 
        Moreover, two sideloaded apps, EvaSpy and TiSpy, obfuscated their source code.
        Connections to trackers like Google Crashlytics and Firebase Analytics frequently failed due to server-side certificate pinning, which could not be bypassed. As a result, we are unable to provide information about the transmission of personally identifiable information to third parties.
        
        We were unable to capture unencrypted traffic for some apps due to factors such as their use of VPNs, custom encryption methods and server-side security checks. However, we successfully obtained unencrypted network traffic for 16 sideloaded apps and 16 in-store apps. For an additional 3 in-store apps, we were only able to collect limited data due to application instability caused by resource mismatches.

        Our analysis revealed that three sideloaded apps -- FlexiSpy, Kidlogger and TheOneSpy -- did not implement Transport Layer Security (TLS)\footnote{\ TLS, used in the HTTPS protocol, encrypts data to ensure secure communication.} encryption. Kidlogger and TheOneSpy transmitted sensitive data in plaintext, whereas FlexiSpy used a custom encryption protocol. Despite employing AES encryption, FlexiSpy's implementation has been shown to be vulnerable to decryption~\cite{Langton2019StalkingStalkerware}, allowing potential attackers to intercept and access sensitive user data.
        In contrast, none of the analysed in-store applications exhibited these issues.

        The network traffic also showed that mSpy transmits sensitive GPS location data on top of device data to Amplitude, a product analytics platform that assists businesses in tracking and analysing user behaviour. 
        We also discovered links between different sideloaded parental control apps. EvaSpy retrieves information hosted on ``172.232.207.34/TiSPY'', while MoniMaster downloads a certificate from ``clevguard.net''. This suggests these apps may share infrastructure or resources, pointing to possible links in their development or distribution.

        \subsection{Stalkerware Indicators of Compromise (IOC)}
        \label{subsec:stalkerwareioc}
        Due to the overlap and misuse of parental control apps and stalkerware (see \autoref{subsec:research-stalkerware}), we decided to match the identified apps to common stalkerware IOC, as described in Q4.
        Our primary source was TinyCheck, a tool designed to detect spyware and stalkerware on mobile devices by analysing network traffic without requiring direct access to the device.
        TinyCheck's IOCs are sourced from Kaspersky’s Global Research and Analysis Team, along with contributions from various researchers and experts, including specialised stalkerware repositories. Their information is free, open-source, regularly updated and used by women's shelters.
        We cross-referenced the domains and corresponding IP addresses identified for each app with these IOCs.        

        Among the 44 identified apps not available on the Google Play Store, $25\%$ were classified as stalkerware by TinyCheck, with one additional app flagged as suspected stalkerware.

        Of the 20 sideloaded apps analysed in this study, eight (or $40\%$) were classified as stalkerware: Cocospy, FlexiSpy, Hoverwatch, iKeyMonitor, mSpy, Spapp Monitoring, TiSpy and XNSPY. Three of them were not detected by Google Play Protect as potentially malicious apps (FlexiSpy, Spapp Monitoring and TiSpy), aligning with past research on the limitations of Google Play Protect~\cite{Hutchinson2019PlayProtect}.

        We further discovered 4 apps currently available on the Google Play Store that are considered stalkerware by the IOC used for this study: Find My Kids: GPS-Tracker, KidsGuard Pro-Phone Monitoring, Life360 and mLite – GPS Location Tracker.

\section{Discussion}
\label{sec:discussion}
    Our analysis of 20 parental control apps from sideloaded sources and 20 from the Google Play Store highlights concerning patterns in the sideloaded parental control market, as evidenced by our discoveries related to the driving research questions.

    Regarding Q1, we found that sideloaded parental control apps appear to offer even more extensive surveillance capabilities, while simultaneously offering fewer specialised parental control features. These apps request a significantly higher number of permissions, including potentially dangerous ones. Particularly alarming is the fact that over half of the analysed parental control apps included features to monitor popular dating apps, which raises serious privacy and safety concerns for children.
    While underage users may find ways to access these platforms, these services have an age requirement of at least 18 years old. The availability of monitoring tools that enable surveillance of these age-restricted apps can hint at the fact that this feature is designed for monitoring the activities of adults. While it can only be speculated, the software providers may endorse surveillance within intimate partner relationships while posing as parental control software.
    When it comes to trackers, we found that Google Play Store apps use significantly more trackers than sideloaded apps. Most of the encountered trackers aim to improve app performance, enhance user experience, or assess marketing effectiveness.
    While obfuscation was not present in Google Play Store apps, most sideloaded apps attempt to hide their activity by pretending to be inconspicuous system-related services. 

    Concerning Q2, we found that parents and children were, in most cases, not sufficiently informed about the sideloaded apps' privacy practices because the policies were often unclear, hard to access, or lacked crucial information.
    In connection with this, we identified multiple suspected violations of the GDPR regarding privacy policies, indicating potential legal and regulatory problems. The absence of privacy policies calls into question the credibility of the analysed app vendors' claims that their apps are intended to provide a safe and secure digital environment for children.

    On the topic of Q3, we found that three sideloaded parental control apps failed to protect confidential user data during transmission by not implementing encryption. We uncovered potential relationships between different parental control vendors and discovered one app sharing sensitive location data with third-party trackers.

    Regarding Q4, we found that $25\%$ of all sideloaded Android parental control apps identified were flagged as stalkerware by TinyCheck. This number increases to $40\%$ for the 20 most popular apps analysed in this study.
    It is important to note that TinyCheck does not offer a comprehensive list of IOCs and does not distinguish between dual-use apps and stalkerware. When we compared our app sample with the IOCs database from Echap~\cite{Echap2024}, a hacker collective that combats sexual violence and provides resources to help organisations and individuals prevent cyber violence, a higher number of apps were flagged as stalkerware. According to their IOCs, $50\%$ of the 44 identified sideloaded parental control apps and $80\%$ of the 20 apps analysed in this study are listed as potential stalkerware. By comparing these results, we noted that TinyCheck appeared to apply stricter criteria for identifying stalkerware, while Echap flagged dual-use apps as well, without distinguishing them from stalkerware. However, since neither source provided a detailed explanation of their IOC selection process or criteria, these differences in classification remain somewhat unclear.

    Below, we discuss the implications of our findings in more detail, assess their adherence to safety-by-design principles and consider the ramifications this has on the legitimacy of the sideloaded parental control software market and legislation.
    

    \subsection{Risks of Sideloaded Parental Control Apps}
      We identified several risk factors associated with the usage of the examined sideloaded parental control apps, which we outline below. 

        \subsubsection{Lack of Meaningful Safeguards Against Misuse}
        We observed that sideloaded parental control apps lacked adequate safeguards, such as regular notifications informing the user of monitoring, which are a necessity to prevent the misuse of parental control apps for malicious purposes~\cite{Khoo2019Fear}. 
        When safeguards are present in sideloaded apps, they depend on the correctness of user-entered data without independently verifying it, meaning the measure is ineffective in preventing misuse.
        In practice, there are user-friendly and privacy-conscious options to ensure apps are not used without someone's consent or awareness, such as periodic reminders that monitoring is active, a permanent icon in the notification bar and accessible activity dashboards. Such safeguards were implemented by 17 in-store parental control apps but were largely absent in sideloaded apps. This issue is exacerbated by the fact that the majority of sideloaded apps obfuscate their name and icon, making them better targets for malicious use.
        
        \subsubsection{Abuse of Special Permissions}
        The second finding concerns the use of the following privileges: display above other apps, device administrator and accessibility. 
        Kidlogger best exemplifies the exploitation of special privileges. The window is immediately minimised when the user navigates to the settings and attempts to open Kidlogger's app information. This repeats when trying to maximise it again, meaning the user can not view essential details, such as granted permissions, or click the ``uninstall'' or ``force stop'' buttons. Combined with the device administrator privilege, which disables the uninstall button on the home screen when a user long-presses the app, the deletion of Kidlogger is challenging. In principle, the device administrator privilege can be revoked in newer Android versions, and the app can be uninstalled afterwards. However, many users, especially minors, are likely unaware of this~\cite{Shan2019DeviceAdmin}. 
        This mechanism has been well-documented in relation to stalkerware~\cite{Yan2019Understanding}, yet it was equally used by both sideloaded and in-store apps.
        Google has made efforts to restrict sideloaded apps from using the accessibility API starting with Android 13, but users can still manually grant this permission to parental control apps if desired~\cite{Titterington2023Android13}.

        \subsubsection{Excessive Monitoring}
        We observed that sideloaded apps, which often offer extensive monitoring features not present in in-store apps, request significantly more permissions than those available on the Google Play Store, including potentially dangerous ones. We found evidence that some of these permissions appear unnecessary for the app's parental control functionalities. The acquisition of excessive permissions enables broad data collection, which is particularly concerning for apps managing children's data. This overprivilege in permissions is a common issue across apps in general~\cite{Wu2013Vendor}. Developers use API methods to implement different app capabilities, which require permissions to work. Due to poor documentation, it is not always clear which permissions are essential, and since missing permissions might lead to app crashes, developers tend to add permissions~\cite{Tahaei2023Permissions} carelessly. Unwarranted permissions can, however, cause problems and make apps insecure and unreliable~\cite{Wang2015Reevaluating}. 
        

        Excessive monitoring was also noted during the app analysis. Sideloaded apps offered concerning capabilities, such as keyloggers or remote camera access, surpassing what is typically provided by in-store parental control apps.
        Hoverwatch, for example, took pictures with the front camera every time the device was unlocked. Other apps let the person monitoring remotely trigger a screen recording, the camera and the microphone. It is up to debate whether parental control apps need such features. It should also be highlighted that these functionalities are common in stalkerware~\cite{Parsons2019Predator}.
        Additionally, more than half of the parental control apps promoted the monitoring of dating apps such as Tinder and Hinge on their websites. As these dating platforms have minimum age requirements of 18 years, the availability of monitoring tools that enable surveillance of such age-restricted apps suggests that these services could be used to track adults' activities rather than children's (i.\,e.\ in the context of intimate partner violence) while also risking outing minors who are, for example, ``closeted'' LGBTQ+ users~\cite{Bouma-Sims2024ClosetedUsers}. 

        \subsubsection{Lack of Encryption}
        Three apps were found without TLS encryption, exposing network traffic to interception and unauthorised access. This negligence, especially involving children's data, poses a serious risk of sensitive information being compromised by malicious actors.
        
        \subsubsection{Circumvention of Google Play Protect Security Safeguards}
        Lastly, our analysis also examined how sideloaded parental control apps circumvent the Google Play Protect security feature. 13 parental control apps were identified as potentially malicious by Google Play Protect. As a result, most apps instruct the user to disable the feature during installation. 
        Furthermore, some apps go a step further, asking users to not only disable the service but also turn off any notifications from it. As a result, users remain unaware that the protective service has been deactivated, exposing their devices to potential risks from other malicious software. This is unacceptable from a parental control perspective, as it subjects children's devices to unnecessary risks. Legitimate sideloaded parental control apps should aim to operate without triggering device protection mechanisms.

        Google is currently launching an initiative to enhance Google Play Protect in selected countries to address malicious apps. According to a statement on their blog, this innovation will significantly improve the detection and blocking of harmful software, having already identified 515,000 new malicious apps~\cite{Google2024Protect}. Whether these improvements will affect the sideloaded parental control market is subject to future research.
        Google has announced an upgrade to Google Play Protect, incorporating a new fraud protection feature designed to block sideloaded apps that request sensitive runtime permissions~\cite{Google2024Protect}. 

    \subsection{Privacy Policy Shortcomings}
        The privacy policy analysis unveiled several gaps in data protection compliance related to the deployment of sideloaded parental control apps. 

        \subsubsection{Missing of App-Specific Privacy Policies}
        Our findings show a lack of comprehensive privacy policies, leaving parents uninformed about data practices, compromising their ability to make informed decisions and raising questions about the vendors' ethical standards and commitment to data privacy, transparency and accountability.
        Half of the analysed services did not include the app in the scope of their privacy policy or offer a separate policy for their app, meaning there was no legal information about the privacy practices of parental control apps in half of the cases. 

        \subsubsection{Absence of Essential Information}
        The remaining ten policies lacked essential details, such as user rights, the types of data collected and which third parties it might be shared with.       
        Therefore, parents using these services cannot inform themselves about the extent of data collection, encompassing their own personal information and, more significantly, that of their child. Besides, both parents and children are in the dark about whether the app is safe to use, as privacy implications and security measures implemented by the services are mostly unclear. These findings echo research by Sun et al.~\cite{Sun2024SmartHome}, who found a similar lack of child safety and privacy information in privacy policies of smart home products explicitly advertised for children. Information about what data a specific product might collect about children and for what purposes was rarely mentioned upfront. 
        Moremen et al.~\cite{Moremen2024Generational} showed that while people across all generations struggle to understand technical terms, misunderstandings and lack of knowledge are particularly acute problems for children.
        Indeed, without explicit privacy guidelines, children using these apps may become unwitting subjects of data collection and monitoring~\cite{Lupton2017Datafied}.
        Minors are left particularly vulnerable if their sensitive information is exposed, such as GPS locations, private messages, browser history, photos and videos.
        Data breaches involving sideloaded parental control apps are not rare, as shown by several breaches in recent years affecting millions of users~\cite{Apostol2024mSpy, Sophos2020KidsGuard, Bitdefender2024PCTattletale}.
  

        \subsubsection{Failure to Obtain Meaningful Consent}
        Our findings also raise critical questions about consent. The GDPR mandates that parental consent be obtained for children under the age of 16~\cite{EuropeanParliament2018GDPR8}. However, Parsons et al.~\cite{Parsons2019Predator} argue that obtaining consent from just one parent is inadequate in joint parenting situations. Each parent with custodial rights must provide informed and verifiable consent to the surveillance of the child. None of the analysed apps implemented measures to ensure this. Parsons et al.\ underscore the importance of mutual consent by addressing situations of intimate partner violence where single-parent consent can lead to harm to the non-consenting partner, who is indirectly subject to the monitoring abilities of the parental control app.
        Some software vendors stated that only the device owner's consent is required for monitoring. However, device ownership extends beyond legal guardianship. This prompts consideration of whether consent given by other relatives (e.\,g.\ step-parents or foster care workers) is sufficient to facilitate child monitoring and how mutual consent provisions could be technically implemented in shared device scenarios.

        \subsubsection{Imbalance in User Rights and Data Autonomy}
        Sideloaded parental control apps present a significant control imbalance regarding user rights. Children lack autonomy over their data, even when the monitoring concludes or they reach adulthood.
        In addition to the frequent omission of user rights in policies, it is crucial to emphasise that, across all cases, only the account creator (typically the assumed parent) possesses the capability to exercise these rights. As a result, the monitored individual is devoid of authority over their own data unless they attain account ownership or initiate direct contact with the software vendor. Even when the use of the parental control app ceases, such as when the child reaches the age of 18, the individual still cannot access or request the deletion of their data. None of the policies addressed this issue or explicitly provided ways for the person whose data is being collected to exercise any authority over their data. As court cases have begun of children suing their parents for breaches of privacy due to parental ``oversharenting'' (i.\,e.\ parents sharing too much or inappropriate information)~\cite{Ong2022Sharenting, Bessant2017Sharenting}, it is plausible that such developments are on the horizon.
        
        \subsubsection{Readability}
        Our results also showed that the readability of all policies ranged from fairly difficult to very difficult.
        Poor readability is a problem among privacy policies in general. A previous study found that the average Flesch–Kincaid readability score for privacy policies of the top one million websites was 39.8~\cite{Libert2018Automated}, which is not significantly better than the average of the analysed parental control policies. Some of the contributing factors to these suboptimal implementations may include communication gaps and adversarial relationships between software developers and privacy experts~\cite{Zou2024CrossContextual}.

        Lastly, it is important to note that even when privacy policies are accessible, research has found that a significant portion of users ($74\%$) simply do not read them~\cite{Obar2020Lie}. Furthermore, even when users do read and understand the policy, they often struggle to translate that information into a clear understanding of the privacy risks they face~\cite{Tsohou2017Consent}.
        This is caused by factors such as the length of policies, the difficulty of comprehending them, vague language and the limitations of real choice~\cite{Borgesius2015Informed}.
        Some scholars thus argue that digital consent is intrinsically flawed~\cite{Carmi2021Feminist}, which further means that parents may not fully grasp what they are agreeing to. These challenges are exacerbated by the incomprehensible and inchoate policies analysed in this work, which often did not adequately address their respective apps.
        
        Regardless of these issues, studying the privacy policies of sideloaded apps remains relevant. Even though users might not read or fully make sense of them, privacy policies serve as a legal framework that outlines the company’s commitments and practices. They provide a baseline for accountability and are usually the only formal source of information about how user data is handled. Moreover, a comprehensive policy allows users to inform themselves, whereas the absence or incompleteness of a policy takes this option away entirely, increasing the risk of uninformed consent.

    \subsection{Safety-by-Design Non-Compliance}
    \label{sec:safetybydesign}
        The studied sideloaded parental control apps severely lack adherence to safety-by-design principles, which give platforms and services guidance to incorporate, evaluate and enhance user safety.
        We used the eSafety Commissioner's recommendations~\cite{eSafetyCommissioner2019Principles} to assess whether the vendors should be trusted with children's data. However, there is an abundance of other efforts underway to centre children's technology design needs, including the UK’s Age Appropriate Design Code~\cite{ICO2020DesignCode} and California’s Age Appropriate Design Code Act~\cite{California2022DesignCode}.
        
        \subsubsection{Service Provider Responsibility}
        The first safety-by-design principle places responsibility on service providers to understand potential online harms and to address them in the design of their systems. Arguably, all analysed app vendors failed this measure, as none implemented effective safeguards, like ongoing or regular status bar notifications showing monitoring is active. We encountered disclaimers on both the vendor's websites as well as in the apps, shifting the responsibility to the app users.
    
        \subsubsection{User Empowerment and Autonomy}
        The second principle focuses on user empowerment and autonomy, stating that the service should align with the best interests of users, particularly the children being monitored. While parental control offers benefits to minors, such as protecting them from inappropriate content or enabling parents to intervene in critical situations such as cyberbullying, common in-store apps can also provide these positive outcomes.
        Conversely, the obfuscation inherent in many sideloaded parental control apps, which often monitor children without their knowledge or consent, undermines their sense of agency and significantly diminishes their autonomy. 
        The notion that excessive monitoring may not serve the best interests of children is also suggested by app reviews: More than half of the Google Play Store reviews for parental control apps are negative~\cite{Akgul2024Reviews}, with ratings from children being notably lower than those from parents~\cite{Ghosh2018Matter}. Recent research even found evidence that overcontrol is linked to brain connectivity patterns tied to threat sensitivity, supporting its classification as childhood trauma~\cite{Carbone2024Overcontrol}. 
        Inadequate security measures also negate this principle's tenets, as well as the lack of comprehensive and complete privacy policies.       
        The second principle recommends the implementation of built-in support functions and feedback loops. 
        Although some parental control apps offered support services for the parent through their website, no such services were available for children or implemented into the apps themselves.
        
        \subsubsection{Transparency and Accountability}
        The third and final principle addresses transparency and accountability. 
        Companies should ensure that user safety policies, terms and conditions, community guidelines and processes about user safety are accessible, easy to find, regularly updated and easily understood.
        Yet, half of the sideloaded parental control apps did not provide privacy policies applicable to the apps, and most policies were not even displayed within the apps.
        
        In conclusion,  most of the analysed sideloaded parental control apps fail to adhere to any of the proposed safety-by-design principles. While past research has shown that these principles may be challenging to implement~\cite{Grace2023ChildDesign}, they provide a solid foundation for vendors which they must be expected to comply with. 
    
    \subsection{Ethical Implications}
    Our research raises several important questions regarding the use of sideloaded parental control apps that warrant careful consideration. The implications of our work, therefore, extend beyond the technical domain, encompassing broader societal concerns.
    
        \subsubsection{Legitimisation of the Sideloaded Parental Control Software Market}
            Considering that more than half of the identified sideloaded parental control apps were matched to stalkerware IOC, it is imperative to understand the danger stalkerware poses. 
            Stalkerware often targets individuals such as partners, ex-partners and family members, granting access to personal and sensitive information. Such an invasion of privacy can have severe emotional, psychological and physical consequences for the affected parties~\cite{Chatterjee2018spyware,Freed2018Stalkers,Parsons2019Predator}.
            Kaspersky's 2023 report identified 31,031 stalkerware victims, an increase from 2022, though the actual number is likely higher as only Kaspersky users were analysed. They emphasise that stalkerware remains a global issue~\cite{Kaspersky2024State}.

            Liu et al.~\cite{Liu2023No} suggested that fines, such as the one imposed against the spyware product StealthGenie in 2014~\cite{OfficeOfPublicAffairs2014}, compelled stalkerware developers trying to avoid similar penalties to market their product for ostensibly lawful purposes, mainly child and employee monitoring.
            Consequently, the market of sideloaded parental control apps appears to be subverted by stalkerware, making it harder to find credible software.
            
\begin{figure}[h]
            \centering
            \includegraphics[width=0.8\linewidth]{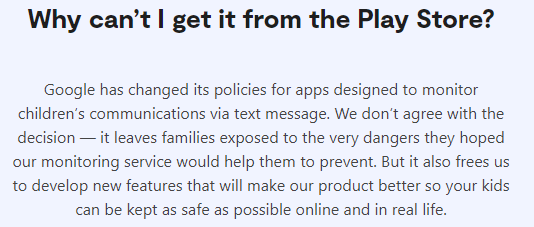}
            \caption{Bark's reasoning behind shifting their app to the sideloaded app market}
            \Description[Image of Bark's statement regarding sideloading]{Google has changed its policies for apps designed to monitor children's communications via text message. We don't agree with the decision it leaves families exposed to the very dangers they hoped our monitoring service would help them to prevent. But it also frees us to develop new features that will make our product better so your kids can be kept as safe as possible online and in real life.}
            \label{img:bark-image}
            \end{figure}

            Despite arguments challenging the legitimacy of the sideloaded parental control market, some vendors have provided reasons why this market is still important.
            Bark used to offer its software on the Google Play Store but has since decided to move to sideloading, citing stricter regulations by the Google Play Store when it comes to monitoring children's communication. They addressed this topic on their website, as seen in \autoref{img:bark-image}. 
            Other vendors point out that sideloaded apps can address situations where children have learned to circumvent or uninstall in-store apps. Sideloaded apps often remain hidden and are more difficult to uninstall, providing parents with a tool to maintain monitoring when they believe it is still necessary.
            Another possible reason for sideloading is to avoid paying service fees and commissions on sales~\cite{Brady2024AppSideloading, Morton2024EntryCompetition}. While this is typically not lucrative, as sideloaded apps are less visible and get fewer customers, the strategy could be more effective if only the children's app is sideloaded while the parent's app remains on the official store. This way, vendors can maintain visibility while sidestepping fees by charging solely for the children's app.

            Aside from the specifics of sideloaded apps, researchers have also argued that parental control apps are not optimal for good parenting. Such top-down monitoring approaches should be replaced with family online safety apps, which focus more on aspects such as children's autonomy, learning and appropriate use of technology~\cite{Ghosh2018Matter} or constructive parent-child interactions such as conversations to shape minors' privacy and security literacy~\cite{Alghythee2024ConversationLiteracy, Theofanopoulou2024Parent-Child, Blinder2024WouldRather, NSPCC2024}.

        \subsubsection{Legal and Regulatory Responses} 
            Policy and legislation response are necessary to protect children's data from being endangered by stalkerware posing as parental control apps. 
            Researchers like Sonia Livingstone, who specialises in children's digital rights, have highlighted the challenges policymakers face in realising children's rights online~\cite{Livingstone2020Rights}.
            She suggests measures such as prioritising digital literacy education for both children and parents and building expertise on digital matters within the child workforce.\footnote{\ Occupations that involve working with or caring for children, such as teaching or childcare.}
            
            Recent developments such as the UK's Online Safety Act (2023) and the proposed USA's Kids Online Safety Act (KOSA) push for such reforms. Still, they primarily focus on obligations for social media platforms and their provisions related to parental control apps and children's data privacy remain heavily contested~\cite{NashFelton2024OSA, CDT2024KOSA}.           
            While existing frameworks, such as the GDPR, set benchmarks to safeguard individuals' data, including children's data, our research uncovered numerous potential GDPR violations with no apparent legal consequences to date. Data protection authorities should, therefore, take a more active role in overseeing the sideloaded app market, particularly those apps claiming to be parental control tools and enforcing compliance with data privacy laws.

           \subsubsection{Advice for App Developers and Parents}
           Our research also shows the need for explicit recommendations for app developers and parents.  The sideloaded software market must adhere to established safety-by-design principles, ensuring their apps have robust privacy protections, clear consent processes and transparent privacy policies from the outset~\cite{Gnanasekaran2023Recommendations}. Developers should also incorporate mechanisms that allow children to exercise their rights over their data once they reach adulthood. 
           Parents should be provided with accessible resources and tools to help them make informed decisions about which parental control apps to install, particularly in light of the DMA, which will open up the availability of third-party apps independent of app stores. Educating parents about the risks of sideloaded apps and promoting the use of safe and child-friendly apps that have undergone rigorous security and privacy checks can help mitigate the risks associated with poorly vetted third-party apps.

        \subsubsection{Misuse of Findings}
            We acknowledge the risk of individuals misusing our work in their search for apps with specific properties or capabilities.
            However, the intent behind this paper was not to advocate or facilitate malicious use. 
            Instead, we aimed to raise awareness in this field and incentivise changes in the practices of the sideloaded parental control market.

        \subsubsection{Coordinated Disclosure}
            In light of the concerns around the potential misuse of these apps and the detected potential security and GDPR breaches, we adhered to the OWASP guide for coordinated disclosure~\cite{OWASP2024Disclosure}. We shared our findings with all 20 sideloaded app vendors in advance, following CERT/CC’s disclosure policy, to give them the opportunity to respond and implement any necessary measures before our results are made public. Due to the absence of critical findings for in-store apps, there was no need for notification in their case. 
            As of publication, only two vendors have responded to our disclosure and we are unaware of any changes being implemented.
            Lastly, the Android Security Team was notified about the seven sideloaded apps with stalkerware traits that were not flagged by Google Play Protect.

\section{Limitations}    
    Several limitations affect our study. First, the Google Play Store’s lack of download number filtering may have led to the exclusion of some popular parental control apps from our selection. Second, issues outlined in \autoref{subseC:traffic} prevented us from analysing all sideloaded apps, limiting our sample.
    Third, static analysis using MobSF and Exodus does not account for dynamic execution, such as runtime permissions or unexecuted code paths involving third-party endpoints.
    Since these tools rely on pre-compiled tracker databases, any trackers absent from these databases may go undetected, leading to potential false negatives. Additionally, the tools may struggle with obfuscation, possibly preventing a complete assessment when heavily obfuscated code is present. 
    Finally, the Flesch-Kincaid readability test, while useful, has limitations, as it only measures quantifiable aspects of readability and assumes uniform reader characteristics, such as cognitive ability and maturity~\cite{Redish2000Readability}.
    Despite these limitations, our findings remain robust, replicable and reproducible, with the potential for addressing these gaps in future research.

\section{Conclusion}
    This study found several privacy, safety and security concerns regarding sideloaded parental control apps.
    We examined the apps of 20 sideloaded and 20 Google Play Store apps as well as their privacy policies.
    We recognised that many issues identified in previous research on in-store parental control apps also apply to the sideloaded versions examined in this study~\cite{Feal2020AngelOD, Okoyomon2019Contradictions}. Some problems are exacerbated in sideloaded apps, including excessive monitoring, inappropriate functionalities and overlap with stalkerware, further compounded by a lack of safeguards.
    Unlike in-store parental control apps, the sideloaded versions provided far more extensive monitoring capabilities, thus requiring more permissions. However, fewer of these sideloaded apps offered specialised parental control functionalities.
    Obfuscation was also more prevalent among the sideloaded apps, while trackers were more common among in-store apps, suggesting a greater focus on user analytics.
    
    The privacy policies for sideloaded apps raised many issues, especially completeness, transparency and availability. Most sideloaded vendors failed to provide children or monitored individuals any means to exercise authority over their personal data and user rights, such as deletion of information. This disregard for children's digital autonomy and privacy is deeply troubling.
    A significant number of sideloaded parental control apps have triggered alerts from stalkerware detection tools, suggesting that certain features of these apps bear a concerning resemblance to those typically associated with surveillance software. 
    
    Moreover, most analysed apps fail to meet basic safety-by-design principles despite handling sensitive and personally identifiable children's data.
    As such, the studied apps cannot be considered ``ethical'' parental control, as the level of surveillance significantly surpasses that of conventional, in-store options. 
    The erosion of children's independence, agency and trust facilitated by sideloaded apps is a grave concern. Sideloaded parental control apps that operate in the shadows without transparency or user consent undermine the foundations of a healthy parent-child relationship built on mutual understanding and respect.

    Based on our findings, we call for greater oversight and regulation of the sideloaded software market to mitigate these products' dangers for children and other vulnerable groups, such as stalking and intimate partner violence victims. Data protection authorities must step in to ensure sideloaded apps adhere to the highest privacy and security standards. Only then can we create a digital landscape that empowers children, preserves their rights and fosters the trust essential for their well-being, protection and development.
    
    We also want to address upcoming changes in the evolving landscape of sideloaded software, particularly with the DMA requiring Apple to open its smartphone ecosystem to sideloaded software. Given these impending shifts, future analyses of sideloaded parental control software and the broader sideloaded software market will be crucial.

    Other valuable research opportunities include interviewing app developers and company executives to gather insights into their views on privacy practices and abuse vectors. Additionally, questioning parents who use sideloaded parental control apps on their motives and awareness of dangers could provide further relevant context. Finally, future research could focus on developing and proposing concrete recommendations for policymakers, app developers and parents based on the identified security and privacy issues in parental control apps.

\begin{acks}
    The authors are grateful to University College London and St.\ Pölten University of Applied Sciences for their institutional support and thank the anonymous peer reviewers for their valuable comments and suggestions. We would also like to thank Ruba Abu-Salma, Jorge Blasco Alís, Kevin Butler, Rahul Chatterjee, George Danezis, Diarmaid Harkin, Etienne Maynier, Steven Murdoch, Julian Nagele, Simon Parkin, Elissa Redmiles, Maria Schett, Laura Shipp, Jun Zhao and Yixin Zou for their helpful input on earlier drafts of this paper as well as the members of the \textit{Gender and Tech Research Lab} for their valuable feedback.

    The authors used AI-based tools (ChatGPT, Grammarly and Lex) to correct typos, grammatical errors and awkward phrasing throughout the paper.

    The second author was supported by a UK Research and Innovation (UKRI) Future Leaders Fellowship (FLF; grant number MR/W009692/1) and the UK Prevention Research Partnership (Violence, Health and Society; grant number MR-VO49879/1).
\end{acks}

\bibliographystyle{ACM-Reference-Format}
\bibliography{sample-base}

\appendix

\section{App Selection Search Terms}
\label{appendix:selectionterms}
 \begin{enumerate}
    \item app to manage kid's phone remotely android
    \item best app to track kid's phone android
    \item best child monitoring apps android
    \item best parental control apps android
    \item child monitoring app android
    \item free child monitoring apps android
    \item free parental control apps android
    \item how to control kid's phone usage android
    \item how to monitor child's online activity android
    \item how to monitor child's phone android
    \item parental control app android
    \item phone monitoring app for kids android
    \item phone monitoring for parents android
    \item remote phone monitoring for parents android
    \item safe phone monitoring for children android
\end{enumerate}

\newpage
\section{Privacy Policies and Trackers}
\label{appendix:tables}
\begin{table}[!ht]
    \centering
    \caption{Parental Control App Privacy Policies}
    \label{table:policies}
    \small
    \begin{tabular}{lcccccc}
        \multicolumn{1}{c}{\rotatebox{90}{\textbf{App Name}}} & 
        \multicolumn{1}{c}{\rotatebox{90}{\textbf{Available}}} & 
        \multicolumn{1}{c}{\rotatebox{90}{\textbf{Applicable}}} & 
        \multicolumn{1}{c}{\rotatebox{90}{\textbf{Scope}}} & 
        \multicolumn{1}{c}{\rotatebox{90}{\textbf{Location}}} & 
        \multicolumn{1}{c}{\rotatebox{90}{\textbf{Flesch-Kincaid Score}}} & 
        \multicolumn{1}{c}{\rotatebox{90}{\textbf{Length in Characters}}} \\ \hline \hline
        Alli360 & \scalebox{0.75}{\ding{108}} & \scalebox{0.75}{\ding{108}} & MA, W & S & 25.47 & 12787 \\ \hline
        AnyControl & \scalebox{0.75}{\ding{108}} & \scalebox{0.75}{\ding{108}} & AS & W & 27.50 & 4080 \\ \hline
        Bark & \scalebox{0.75}{\ding{108}} & \scalebox{0.75}{\ding{108}} & A, W & W & 30.55 & 26104 \\ \hline
        Chyldmonitor & \scalebox{0.75}{\ding{108}} & ~ & W & W & ~ & ~ \\ \hline
        Cocospy & \scalebox{0.75}{\ding{108}} & ~ & W & A & ~ & ~ \\ \hline
        ESET Parental Control & \scalebox{0.75}{\ding{108}} & \scalebox{0.75}{\ding{108}} & AS & S & 34.16 & 12305 \\ \hline
        EvaSpy & \scalebox{0.75}{\ding{108}} & ~ & W & A & ~ & ~ \\ \hline
        FamilyTime Jr. & \scalebox{0.75}{\ding{108}} & \scalebox{0.75}{\ding{108}} & A & A & 29.33 & 10915 \\ \hline
        FamiSafe Kids & \scalebox{0.75}{\ding{108}} & \scalebox{0.75}{\ding{108}} & AS & S & 20.51 & 23213 \\ \hline
        FlexiSpy & \scalebox{0.75}{\ding{108}} & ~ & W & W & ~ & ~ \\ \hline
        Hoverwatch & \scalebox{0.75}{\ding{108}} & \scalebox{0.75}{\ding{108}} & AS & W & 41.73 & 17877 \\ \hline
        iKeyMonitor & \scalebox{0.75}{\ding{108}} & ~ & W & W & ~ & ~ \\ \hline
        iSharing & \scalebox{0.75}{\ding{108}} & \scalebox{0.75}{\ding{108}} & A, W & S & 22.84 & 20972 \\ \hline
        Kaspersky SafeKids & \scalebox{0.75}{\ding{108}} & \scalebox{0.75}{\ding{108}} & AS & S & 18.06 & 34358 \\ \hline
        KidControl & ~ & ~ & ~ & ~ & ~ & ~ \\ \hline
        Kidlogger & \scalebox{0.75}{\ding{108}} & ~ & W & W & ~ & ~ \\ \hline
        Kids Place & \scalebox{0.75}{\ding{108}} & \scalebox{0.75}{\ding{108}} & A, W & S & 32.22 & 9954 \\ \hline
        Kidslox & \scalebox{0.75}{\ding{108}} & \scalebox{0.75}{\ding{108}} & AS & S & 40.83 & 9110 \\ \hline
        Kidstracker & \scalebox{0.75}{\ding{108}} & \scalebox{0.75}{\ding{108}} & AS & W & 27.88 & 9736 \\ \hline
        Kidsy & \scalebox{0.75}{\ding{108}} & \scalebox{0.75}{\ding{108}} & MA & S & 44.61 & 37385 \\ \hline
        Kroha & \scalebox{0.75}{\ding{108}} & \scalebox{0.75}{\ding{108}} & A, W & S & 28.22 & 22687 \\ \hline
        Life360 & \scalebox{0.75}{\ding{108}} & \scalebox{0.75}{\ding{108}} & AS & S & 27.87 & 57578 \\ \hline
        Microsoft Family Safety & \scalebox{0.75}{\ding{108}} & \scalebox{0.75}{\ding{108}} & AS & S & 30.12 & 31399 \\ \hline
        MMGuardian & \scalebox{0.75}{\ding{108}} & \scalebox{0.75}{\ding{108}} & AS & A & 25.45 & 24487 \\ \hline
        MobileFence & \scalebox{0.75}{\ding{108}} & \scalebox{0.75}{\ding{108}} & A, W & S & 34.91 & 3717 \\ \hline
        MoniMaster & \scalebox{0.75}{\ding{108}} & \scalebox{0.75}{\ding{108}} & A, W & A & 18.53 & 11675 \\ \hline
        mSpy & \scalebox{0.75}{\ding{108}} & ~ & W & W & ~ & ~ \\ \hline
        Pingo & \scalebox{0.75}{\ding{108}} & \scalebox{0.75}{\ding{108}} & AS & S & 38.85 & 36448 \\ \hline
        Qustodio & \scalebox{0.75}{\ding{108}} & \scalebox{0.75}{\ding{108}} & AS & S & 28.26 & 41844 \\ \hline
        Spapp Monitoring & \scalebox{0.75}{\ding{108}} & \scalebox{0.75}{\ding{108}} & A, W & W & 30.36 & 27996 \\ \hline
        Spy Phone Labs Phone Tracker & \scalebox{0.75}{\ding{108}} & \scalebox{0.75}{\ding{108}} & A, W & A & 51.84 & 6755 \\ \hline
        SPY24 & \scalebox{0.75}{\ding{108}} & \scalebox{0.75}{\ding{108}} & A, W & W & 37.38 & 10019 \\ \hline
        SPYX & \scalebox{0.75}{\ding{108}} & ~ & W & A & ~ & ~ \\ \hline
        ST Kids App & \scalebox{0.75}{\ding{108}} & \scalebox{0.75}{\ding{108}} & AS & S & 30.67 & 16952 \\ \hline
        TheOneSpy & \scalebox{0.75}{\ding{108}} & ~ & W & W & ~ & ~ \\ \hline
        Tigrow! & \scalebox{0.75}{\ding{108}} & \scalebox{0.75}{\ding{108}} & A, W & S & 32.05 & 19728 \\ \hline
        TiSpy & \scalebox{0.75}{\ding{108}} & ~ & W & A & ~ & ~ \\ \hline
        uMobix & \scalebox{0.75}{\ding{108}} & \scalebox{0.75}{\ding{108}} & A, W & W & 54.95 & 19454 \\ \hline
        WebWatcher & \scalebox{0.75}{\ding{108}} & U & U & W & 28.15 & 13007 \\ \hline
        XNSPY & \scalebox{0.75}{\ding{108}} & \scalebox{0.75}{\ding{108}} & AS & W & 30.27 & 12365 \\ \hline \hline
        \multicolumn{7}{c}{\small \parbox{0.9\linewidth}{\textbf{Abbreviations:} A = App, AS = All Services, MA = Multiple Apps, \\ S = Google Play Store, U = Unclear, W = Website}} \\
    \end{tabular}
\end{table}

\label{appendix:trackers}
\begin{figure*}[!ht]
    \captionsetup{type=table}
    \centering
    \caption{Trackers Used by Parental Control Apps}
    \label{table:tracker}
    \small
    \resizebox{\textwidth}{!}{
    \begin{tabular}{l|c|c|c|c|c|c|c|c|c|c|c|c|c|c|c|c|c|c|c|c|c|c|c|c|c|c|c|c|c|c}
        \multicolumn{1}{c}{\rotatebox{90}{\textbf{App Name}}} & \multicolumn{1}{c}{\rotatebox{90}{\textbf{ACRA}}} & \multicolumn{1}{c}{\rotatebox{90}{\textbf{Adjust}}} & \multicolumn{1}{c}{\rotatebox{90}{\textbf{Amplitude}}} & \multicolumn{1}{c}{\rotatebox{90}{\textbf{AppsFlyer}}} & \multicolumn{1}{c}{\rotatebox{90}{\textbf{Branch}}} & \multicolumn{1}{c}{\rotatebox{90}{\textbf{Braze}}} & \multicolumn{1}{c}{\rotatebox{90}{\textbf{Didomi}}} & \multicolumn{1}{c}{\rotatebox{90}{\textbf{Facebook Analytics}}} & \multicolumn{1}{c}{\rotatebox{90}{\textbf{Facebook Login}}} & \multicolumn{1}{c}{\rotatebox{90}{\textbf{Facebook Share}}} & \multicolumn{1}{c}{\rotatebox{90}{\textbf{Flurry}}} & \multicolumn{1}{c}{\rotatebox{90}{\textbf{Google Admob}}} & \multicolumn{1}{c}{\rotatebox{90}{\textbf{Google Analytics}}} & \multicolumn{1}{c}{\rotatebox{90}{\textbf{Google CrashLytics}}} & \multicolumn{1}{c}{\rotatebox{90}{\textbf{\parbox{3cm}{Google Firebase \\ Analytics}}}} & \multicolumn{1}{c}{\rotatebox{90}{\textbf{Google Tag Manager}}} & \multicolumn{1}{c}{\rotatebox{90}{\textbf{\parbox{3cm}{Huawei Mobile \\ Services Core}}}} & \multicolumn{1}{c}{\rotatebox{90}{\textbf{Huq Sourcekit}}} & \multicolumn{1}{c}{\rotatebox{90}{\textbf{Instabug}}} & \multicolumn{1}{c}{\rotatebox{90}{\textbf{\parbox{3cm}{Microsoft Visual Studio \\ App Center Analytics}}}} & \multicolumn{1}{c}{\rotatebox{90}{\textbf{\parbox{3cm}{Microsoft Visual Studio \\ App Center Crashes}}}} & \multicolumn{1}{c}{\rotatebox{90}{\textbf{OneSignal}}} & \multicolumn{1}{c}{\rotatebox{90}{\textbf{Pusher}}} & \multicolumn{1}{c}{\rotatebox{90}{\textbf{Segment}}} & \multicolumn{1}{c}{\rotatebox{90}{\textbf{Sensors Analytics}}} & \multicolumn{1}{c}{\rotatebox{90}{\textbf{Sentiance}}} & \multicolumn{1}{c}{\rotatebox{90}{\textbf{Sentry}}} & \multicolumn{1}{c}{\rotatebox{90}{\textbf{Singular}}} & \multicolumn{1}{c}{\rotatebox{90}{\textbf{Umeng Analytics}}}    
        \\ \hline \hline
        Alli360 & ~ & ~ & ~ & \scalebox{0.75}{\ding{108}} & ~ & ~ & ~ & ~ & ~ & ~ & ~ & ~ & ~ & \scalebox{0.75}{\ding{108}} & \scalebox{0.75}{\ding{108}} & ~ & \scalebox{0.75}{\ding{108}} & ~ & ~ & ~ & ~ & ~ & ~ & ~ & ~ & ~ & ~ & ~ & ~ \\ \hline
        AnyControl & ~ & ~ & ~ & ~ & ~ & ~ & ~ & ~ & ~ & ~ & ~ & ~ & ~ & \scalebox{0.75}{\ding{108}} & \scalebox{0.75}{\ding{108}} & ~ & ~ & ~ & ~ & ~ & ~ & ~ & ~ & ~ & ~ & ~ & ~ & ~ & ~ \\ \hline
        Bark & ~ & ~ & ~ & ~ & ~ & ~ & ~ & ~ & ~ & ~ & ~ & ~ & ~ & ~ & \scalebox{0.75}{\ding{108}} & ~ & \scalebox{0.75}{\ding{108}} & ~ & ~ & ~ & ~ & \scalebox{0.75}{\ding{108}} & ~ & ~ & ~ & ~ & \scalebox{0.75}{\ding{108}} & ~ & ~ \\ \hline
        Chyldmonitor & ~ & ~ & ~ & ~ & ~ & ~ & ~ & ~ & ~ & ~ & ~ & ~ & ~ & ~ & \scalebox{0.75}{\ding{108}} & ~ & ~ & ~ & ~ & ~ & ~ & ~ & ~ & ~ & ~ & ~ & ~ & ~ & ~ \\ \hline
        Cocospy & ~ & ~ & ~ & ~ & ~ & ~ & ~ & ~ & ~ & ~ & ~ & ~ & ~ & ~ & ~ & ~ & ~ & ~ & ~ & ~ & ~ & ~ & ~ & ~ & ~ & ~ & ~ & ~ & \scalebox{0.75}{\ding{108}} \\ \hline
        ESET Parental Control & ~ & ~ & ~ & ~ & ~ & ~ & ~ & ~ & ~ & ~ & ~ & ~ & ~ & \scalebox{0.75}{\ding{108}} & ~ & ~ & ~ & ~ & ~ & ~ & ~ & ~ & ~ & ~ & ~ & ~ & ~ & ~ & ~ \\ \hline
        EvaSpy & ~ & ~ & ~ & ~ & ~ & ~ & ~ & ~ & ~ & ~ & ~ & ~ & ~ & ~ & ~ & ~ & ~ & ~ & ~ & ~ & ~ & ~ & ~ & ~ & ~ & ~ & ~ & ~ & ~ \\ \hline
        FamilyTime Jr. & ~ & ~ & ~ & ~ & ~ & ~ & ~ & ~ & ~ & ~ & ~ & ~ & ~ & \scalebox{0.75}{\ding{108}} & \scalebox{0.75}{\ding{108}} & ~ & ~ & ~ & ~ & ~ & ~ & ~ & \scalebox{0.75}{\ding{108}} & ~ & ~ & ~ & ~ & ~ & ~ \\ \hline
        FamiSafe Kids & ~ & \scalebox{0.75}{\ding{108}} & ~ & \scalebox{0.75}{\ding{108}} & ~ & ~ & ~ & \scalebox{0.75}{\ding{108}} & \scalebox{0.75}{\ding{108}} & \scalebox{0.75}{\ding{108}} & ~ & ~ & ~ & \scalebox{0.75}{\ding{108}} & \scalebox{0.75}{\ding{108}} & ~ & ~ & ~ & ~ & ~ & ~ & ~ & ~ & ~ & \scalebox{0.75}{\ding{108}} & ~ & ~ & ~ & ~ \\ \hline
        FlexiSpy & ~ & ~ & ~ & ~ & ~ & ~ & ~ & ~ & ~ & ~ & ~ & ~ & ~ & ~ & \scalebox{0.75}{\ding{108}} & ~ & ~ & ~ & ~ & ~ & ~ & ~ & ~ & ~ & ~ & ~ & ~ & ~ & ~ \\ \hline
        Hoverwatch & ~ & ~ & ~ & ~ & ~ & ~ & ~ & ~ & ~ & ~ & ~ & ~ & ~ & ~ & ~ & ~ & ~ & ~ & ~ & ~ & ~ & ~ & ~ & ~ & ~ & ~ & ~ & ~ & ~ \\ \hline
        iKeyMonitor & \scalebox{0.75}{\ding{108}} & ~ & ~ & ~ & ~ & ~ & ~ & ~ & ~ & ~ & ~ & ~ & ~ & \scalebox{0.75}{\ding{108}} & \scalebox{0.75}{\ding{108}} & ~ & ~ & ~ & ~ & ~ & ~ & ~ & ~ & ~ & ~ & ~ & ~ & ~ & ~ \\ \hline
        iSharing & ~ & ~ & ~ & \scalebox{0.75}{\ding{108}} & ~ & ~ & \scalebox{0.75}{\ding{108}} & ~ & \scalebox{0.75}{\ding{108}} & \scalebox{0.75}{\ding{108}} & ~ & \scalebox{0.75}{\ding{108}} & ~ & \scalebox{0.75}{\ding{108}} & \scalebox{0.75}{\ding{108}} & ~ & \scalebox{0.75}{\ding{108}} & \scalebox{0.75}{\ding{108}} & ~ & ~ & ~ & ~ & ~ & ~ & ~ & ~ & ~ & ~ & ~ \\ \hline
        Kaspersky SafeKids & ~ & ~ & ~ & \scalebox{0.75}{\ding{108}} & ~ & ~ & ~ & ~ & ~ & ~ & ~ & ~ & \scalebox{0.75}{\ding{108}} & \scalebox{0.75}{\ding{108}} & \scalebox{0.75}{\ding{108}} & \scalebox{0.75}{\ding{108}} & \scalebox{0.75}{\ding{108}} & ~ & ~ & ~ & ~ & ~ & ~ & ~ & ~ & ~ & ~ & ~ & ~ \\ \hline
        KidControl & ~ & ~ & ~ & ~ & ~ & ~ & ~ & ~ & ~ & ~ & ~ & \scalebox{0.75}{\ding{108}} & ~ & \scalebox{0.75}{\ding{108}} & \scalebox{0.75}{\ding{108}} & ~ & \scalebox{0.75}{\ding{108}} & ~ & ~ & ~ & ~ & ~ & ~ & ~ & ~ & ~ & ~ & ~ & ~ \\ \hline
        Kidlogger & ~ & ~ & ~ & ~ & ~ & ~ & ~ & ~ & ~ & ~ & ~ & ~ & ~ & ~ & ~ & ~ & ~ & ~ & ~ & ~ & ~ & ~ & ~ & ~ & ~ & ~ & ~ & ~ & ~ \\ \hline
        Kids Place & ~ & ~ & ~ & ~ & ~ & ~ & ~ & ~ & ~ & ~ & ~ & \scalebox{0.75}{\ding{108}} & \scalebox{0.75}{\ding{108}} & \scalebox{0.75}{\ding{108}} & \scalebox{0.75}{\ding{108}} & ~ & ~ & ~ & ~ & ~ & ~ & ~ & ~ & ~ & ~ & ~ & ~ & ~ & ~ \\ \hline
        Kidslox & ~ & ~ & \scalebox{0.75}{\ding{108}} & ~ & ~ & ~ & ~ & ~ & \scalebox{0.75}{\ding{108}} & ~ & ~ & ~ & \scalebox{0.75}{\ding{108}} & \scalebox{0.75}{\ding{108}} & \scalebox{0.75}{\ding{108}} & \scalebox{0.75}{\ding{108}} & ~ & ~ & ~ & ~ & ~ & ~ & ~ & ~ & ~ & ~ & ~ & \scalebox{0.75}{\ding{108}} & ~ \\ \hline
        Kidstracker & ~ & ~ & ~ & ~ & ~ & ~ & ~ & ~ & ~ & ~ & ~ & ~ & ~ & \scalebox{0.75}{\ding{108}} & ~ & ~ & ~ & ~ & ~ & ~ & ~ & ~ & ~ & ~ & ~ & ~ & ~ & ~ & ~ \\ \hline
        Kidsy & ~ & ~ & ~ & \scalebox{0.75}{\ding{108}} & ~ & ~ & ~ & ~ & ~ & ~ & ~ & ~ & ~ & \scalebox{0.75}{\ding{108}} & \scalebox{0.75}{\ding{108}} & ~ & ~ & ~ & ~ & ~ & ~ & ~ & ~ & ~ & ~ & ~ & ~ & ~ & ~ \\ \hline
        Kroha & ~ & ~ & ~ & \scalebox{0.75}{\ding{108}} & ~ & ~ & ~ & ~ & ~ & ~ & ~ & ~ & ~ & \scalebox{0.75}{\ding{108}} & \scalebox{0.75}{\ding{108}} & ~ & ~ & ~ & ~ & ~ & ~ & ~ & ~ & ~ & ~ & ~ & ~ & ~ & ~ \\ \hline
        Life360 & ~ & ~ & ~ & \scalebox{0.75}{\ding{108}} & \scalebox{0.75}{\ding{108}} & ~ & ~ & ~ & ~ & ~ & ~ & \scalebox{0.75}{\ding{108}} & ~ & \scalebox{0.75}{\ding{108}} & \scalebox{0.75}{\ding{108}} & ~ & ~ & ~ & ~ & ~ & ~ & ~ & ~ & ~ & ~ & ~ & ~ & ~ & ~ \\ \hline
        Microsoft Family Safety & ~ & ~ & ~ & ~ & ~ & ~ & ~ & ~ & ~ & ~ & ~ & ~ & ~ & ~ & ~ & ~ & ~ & ~ & ~ & \scalebox{0.75}{\ding{108}} & \scalebox{0.75}{\ding{108}} & ~ & ~ & ~ & ~ & \scalebox{0.75}{\ding{108}} & ~ & ~ & ~ \\ \hline
        MMGuardian & ~ & ~ & ~ & ~ & ~ & ~ & ~ & \scalebox{0.75}{\ding{108}} & \scalebox{0.75}{\ding{108}} & ~ & ~ & \scalebox{0.75}{\ding{108}} & \scalebox{0.75}{\ding{108}} & \scalebox{0.75}{\ding{108}} & \scalebox{0.75}{\ding{108}} & \scalebox{0.75}{\ding{108}} & ~ & ~ & ~ & ~ & ~ & ~ & ~ & ~ & ~ & ~ & ~ & ~ & ~ \\ \hline
        MobileFence & ~ & ~ & ~ & ~ & ~ & ~ & ~ & ~ & ~ & ~ & ~ & ~ & ~ & \scalebox{0.75}{\ding{108}} & \scalebox{0.75}{\ding{108}} & ~ & ~ & ~ & ~ & ~ & ~ & ~ & ~ & ~ & ~ & ~ & ~ & ~ & ~ \\ \hline
        MoniMaster & ~ & ~ & ~ & ~ & ~ & ~ & ~ & ~ & ~ & ~ & ~ & ~ & ~ & \scalebox{0.75}{\ding{108}} & \scalebox{0.75}{\ding{108}} & ~ & ~ & ~ & ~ & ~ & ~ & ~ & ~ & ~ & ~ & ~ & ~ & ~ & ~ \\ \hline
        mSpy & ~ & ~ & \scalebox{0.75}{\ding{108}} & \scalebox{0.75}{\ding{108}} & ~ & ~ & ~ & ~ & ~ & ~ & ~ & ~ & ~ & \scalebox{0.75}{\ding{108}} & \scalebox{0.75}{\ding{108}} & ~ & ~ & ~ & ~ & ~ & ~ & ~ & ~ & ~ & ~ & ~ & ~ & ~ & ~ \\ \hline
        Pingo & ~ & ~ & ~ & \scalebox{0.75}{\ding{108}} & ~ & ~ & ~ & ~ & ~ & ~ & ~ & ~ & ~ & \scalebox{0.75}{\ding{108}} & \scalebox{0.75}{\ding{108}} & ~ & ~ & ~ & ~ & ~ & ~ & ~ & ~ & ~ & ~ & ~ & ~ & ~ & ~ \\ \hline
        Qustodio & ~ & ~ & ~ & ~ & ~ & ~ & ~ & ~ & ~ & ~ & ~ & ~ & ~ & \scalebox{0.75}{\ding{108}} & \scalebox{0.75}{\ding{108}} & ~ & ~ & ~ & ~ & ~ & ~ & ~ & ~ & \scalebox{0.75}{\ding{108}} & ~ & ~ & ~ & ~ & ~ \\ \hline
        Spapp Monitoring & ~ & ~ & ~ & ~ & ~ & ~ & ~ & ~ & ~ & ~ & ~ & ~ & ~ & ~ & ~ & ~ & ~ & ~ & ~ & ~ & ~ & ~ & ~ & ~ & ~ & ~ & ~ & ~ & ~ \\ \hline
        Spy Phone Labs Phone Tracker & ~ & ~ & ~ & ~ & ~ & ~ & ~ & ~ & ~ & ~ & ~ & ~ & ~ & ~ & ~ & ~ & ~ & ~ & ~ & ~ & ~ & ~ & ~ & ~ & ~ & ~ & ~ & ~ & ~ \\ \hline
        SPY24 & ~ & ~ & ~ & ~ & ~ & ~ & ~ & ~ & ~ & ~ & ~ & ~ & ~ & ~ & ~ & ~ & ~ & ~ & \scalebox{0.75}{\ding{108}} & ~ & ~ & ~ & ~ & ~ & ~ & ~ & ~ & ~ & ~ \\ \hline
        SPYX & ~ & ~ & ~ & ~ & ~ & ~ & ~ & ~ & ~ & ~ & ~ & ~ & ~ & ~ & ~ & ~ & ~ & ~ & ~ & ~ & ~ & ~ & ~ & ~ & ~ & ~ & ~ & ~ & ~ \\ \hline
        ST Kids App & ~ & ~ & ~ & ~ & ~ & ~ & ~ & \scalebox{0.75}{\ding{108}} & \scalebox{0.75}{\ding{108}} & \scalebox{0.75}{\ding{108}} & ~ & ~ & \scalebox{0.75}{\ding{108}} & \scalebox{0.75}{\ding{108}} & \scalebox{0.75}{\ding{108}} & \scalebox{0.75}{\ding{108}} & ~ & ~ & ~ & ~ & ~ & ~ & ~ & ~ & ~ & ~ & ~ & ~ & ~ \\ \hline
        TheOneSpy & ~ & ~ & ~ & ~ & ~ & ~ & ~ & ~ & ~ & ~ & ~ & ~ & ~ & \scalebox{0.75}{\ding{108}} & \scalebox{0.75}{\ding{108}} & ~ & ~ & ~ & ~ & ~ & ~ & ~ & ~ & ~ & ~ & ~ & ~ & ~ & ~ \\ \hline
        Tigrow! & ~ & ~ & ~ & ~ & ~ & ~ & ~ & ~ & ~ & ~ & ~ & ~ & ~ & \scalebox{0.75}{\ding{108}} & ~ & ~ & ~ & ~ & ~ & ~ & ~ & ~ & ~ & ~ & ~ & ~ & ~ & ~ & ~ \\ \hline
        TiSpy & ~ & ~ & ~ & ~ & ~ & \scalebox{0.75}{\ding{108}} & ~ & ~ & ~ & ~ & \scalebox{0.75}{\ding{108}} & ~ & ~ & ~ & ~ & ~ & ~ & ~ & ~ & ~ & ~ & ~ & ~ & ~ & ~ & ~ & ~ & ~ & ~ \\ \hline
        uMobix & ~ & ~ & \scalebox{0.75}{\ding{108}} & ~ & ~ & ~ & ~ & ~ & ~ & ~ & ~ & \scalebox{0.75}{\ding{108}} & ~ & \scalebox{0.75}{\ding{108}} & \scalebox{0.75}{\ding{108}} & ~ & ~ & ~ & ~ & ~ & ~ & ~ & ~ & ~ & ~ & ~ & ~ & ~ & ~ \\ \hline
        WebWatcher & ~ & ~ & ~ & ~ & ~ & ~ & ~ & ~ & ~ & ~ & ~ & ~ & ~ & \scalebox{0.75}{\ding{108}} & \scalebox{0.75}{\ding{108}} & ~ & ~ & ~ & ~ & ~ & ~ & ~ & ~ & ~ & ~ & ~ & ~ & ~ & ~ \\ \hline
        XNSPY & ~ & ~ & ~ & ~ & ~ & ~ & ~ & ~ & ~ & ~ & ~ & \scalebox{0.75}{\ding{108}} & \scalebox{0.75}{\ding{108}} & \scalebox{0.75}{\ding{108}} & \scalebox{0.75}{\ding{108}} & \scalebox{0.75}{\ding{108}} & ~ & ~ & ~ & ~ & ~ & ~ & ~ & ~ & ~ & ~ & ~ & ~ & ~ \\
        \hline \hline
    \end{tabular}
    }
    \Description[Table containing information about the privacy policies of the analysed apps.]{}
\end{figure*}

\end{document}